%% file: main.tex
\documentclass[conference]{IEEEtran}
\usepackage{algorithm}
\usepackage{algorithmic}
\usepackage{amsmath,amssymb,amsfonts,amsthm}
\usepackage{ascmac}
\usepackage{bm}
\usepackage{cite}
\usepackage{footnote}
\usepackage{graphicx}
\usepackage{mathrsfs}
\usepackage{multirow}
\usepackage[subrefformat=parens]{subcaption}
\usepackage{textcomp}
\usepackage{threeparttable}
\usepackage[normalem]{ulem}
\usepackage{url}
\usepackage{wasysym}
\usepackage{xcolor}
\usepackage{xspace}

\newcommand{\tNAME}{\texttt{NHTD-GL}\xspace}

\newcommand{\vecd}{\bm{d}}
\newcommand{\vech}{\bm{h}}
\newcommand{\vechl}{\bm{h}^{(l)}}
\newcommand{\vechL}{\bm{h}^{(L)}}
\newcommand{\vecm}{\bm{m}}

\newcommand{\vecx}{\bm{x}}
\newcommand{\vecy}{\bm{y}}
\newcommand{\vecz}{\bm{z}}

\newcommand{\matWl}{\bm{W}^{(l)}}

\newcommand{\funcFg}{\mathcal{F}_{\mathsf{enc}}}
\newcommand{\funcfc}{f_{\mathsf{pred}}}
\newcommand{\funcfg}{f_{\mathsf{enc}}}
\newcommand{\funcfgwith}[1]{f_{\mathsf{enc}; {#1}}}
\newcommand{\funcH}{\mathcal{H}}
\newcommand{\funcL}{\mathcal{L}}
\newcommand{\funcN}{\mathcal{N}}
\newcommand{\funcNk}{\mathcal{N}_{k}}
\newcommand{\funcO}{\mathcal{O}}
\newcommand{\funcMLP}{\mathrm{MLP}}
\newcommand{\funcACT}{\psi}
\newcommand{\funcType}{\mathsf{type}}
\newcommand{\funcBehavior}{\mathsf{nb}}
\newcommand{\funcFB}{\mathsf{fb}}
\newcommand{\setVt}{V_\mathsf{t}}
\newcommand{\setVn}{V_\mathsf{n}}
\newcommand{\Geaug}{G_{\mathsf{EAUG}}}
\newcommand{\setB}{\mathcal{B}}

\newcommand{\tCircuitLevel}{circuit-wise }
\newcommand{\tNodeLevel}{node-wise }
\newcommand{\TNodeLevel}{Node-wise }
\newcommand{\TTNodeLevel}{Node-Wise }

\newcommand{\tMotivationA}{\textit{Motivation 1}\xspace}
\newcommand{\tMotivationB}{\textit{Motivation 2}\xspace}
\newcommand{\tMotivationC}{\textit{Motivation 3}\xspace}

\newcommand{\ResFigExt}{pdf}

\newcommand{\mulrowchk}{\multirow{2}{*}{\checkmark}}

\newcommand{\circled}[1]{{\ooalign{{\large \CIRCLE}\crcr{\hss\textcolor{white}{\footnotesize #1}\hss}}}}
\newcommand{\circleedged}[1]{{\ooalign{{\large \Circle}\crcr{\hss\raisebox{0.2mm}{\small #1}\hss}}}}

\newtheorem{definition}{Definition}
\newtheorem{proposition}{Proposition}
\newtheorem{theorem}{Theorem}

\renewcommand{\algorithmiccomment}[1]{\bgroup\hfill//~#1\egroup}

\newcommand{\pagelimitmarker}[1]{~\\ {\textcolor{red}{\ifthenelse{\thepage>#1}{\Huge Exceeding the page limit}{\huge Within the page limit}}}~\\ {\huge{\textcolor{red}{~~Page Limit\,\,\,\,\, = #1}}}~\\ {\huge{\textcolor{red}{~~Current Page = $\thepage$}}}}

\newif\ifcommentson\commentsontrue
\ifcommentson
\newcommand{\commentsize}[0]{.45\textwidth}
\newcommand{\commentKTH}[1]{\begin{center}
\parbox{\commentsize}{\textbf{\textcolor{black}{KTH.}} \textcolor{red}{#1} }\end{center}}
\newcommand{\commentKY}[1]{\begin{center}
\parbox{\commentsize}{\textbf{\textcolor{black}{KY.}} \textcolor{red}{#1} }\end{center}}
\newcommand{\commentSH}[1]{\begin{center}
\parbox{\commentsize}{\textbf{\textcolor{black}{SH.}} \textcolor{red}{#1} }\end{center}}
\newcommand{\commentKF}[1]{\begin{center} \parbox{\commentsize}{\textbf{\textcolor{black}{KF.}} \textcolor{red}{#1} }\end{center}}
\newcommand{\commentNT}[1]{\begin{center} \parbox{\commentsize}{\textbf{\textcolor{black}{NT.}} \textcolor{red}{#1} }\end{center}}
\newcommand{\commentKH}[1]{\begin{center} \parbox{\commentsize}{\textbf{\textcolor{black}{KH.}} \textcolor{red}{#1 }}\end{center}}
\else
\newcommand{\commentKTH}[1]{}
\newcommand{\commentKY}[1]{}
\newcommand{\commentSH}[1]{}
\newcommand{\commentKF}[1]{}
\newcommand{\commentNT}[1]{}
\newcommand{\commentKH}[1]{}
\fi

\newif\ifconferenceon\conferenceontrue
\ifconferenceon
\newcommand{\arxiv}[1]{}
\newcommand{\conference}[1]{#1}
\else
\newcommand{\arxiv}[1]{#1}
\newcommand{\conference}[1]{}
\fi

\allowdisplaybreaks[1]

\begin{document}

\title{Node-wise Hardware Trojan Detection\\Based on Graph Learning}

\author{
    \IEEEauthorblockN{
        Kento Hasegawa\IEEEauthorrefmark{1}\textsuperscript{\textsection},
        Kazuki Yamashita\IEEEauthorrefmark{2}\textsuperscript{\textsection},
        Seira Hidano\IEEEauthorrefmark{1},
        Kazuhide Fukushima\IEEEauthorrefmark{1} \\
        Kazuo Hashimoto\IEEEauthorrefmark{2}, and
        Nozomu Togawa\IEEEauthorrefmark{2}
    }
    \IEEEauthorblockA{
        \IEEEauthorrefmark{1}KDDI Research, Inc.,
        \IEEEauthorrefmark{2}Waseda University
    }
}

\maketitle

\begingroup\renewcommand\thefootnote{\textsection}
\footnotetext{Equal contribution}
\endgroup

\begin{abstract}
	\input{tex/0_abstract.tex}

\end{abstract}


\input{tex/1_introduction.tex}

\input{tex/2_backgrounds.tex}

\input{tex/3_motivation.tex}

\input{tex/4_ht_features.tex}

\input{tex/5_proposed.tex}

\input{tex/6_evaluation.tex}

\input{tex/7_conclusion.tex}


\bibliographystyle{bib/IEEEtran}
\bibliography{bib/citation}

\input{tex/8_appendix.tex}

\end{document}

%% file: tex/0_abstract.tex
In the fourth industrial revolution, securing the protection of the supply chain has become an ever-growing concern. One such cyber threat is a hardware Trojan (HT), a malicious modification to an IC. HTs are often identified in the hardware manufacturing process, but should be removed earlier, when the design is being specified. Machine learning-based HT detection in gate-level netlists is an efficient approach to identify HTs at the early stage. However, feature-based modeling has limitations in discovering an appropriate set of HT features. We thus propose \tNAME in this paper, a novel node-wise HT detection method based on graph learning~(GL). Given the formal analysis of HT features obtained from domain knowledge, \tNAME bridges the gap between graph representation learning and feature-based HT detection. The experimental results demonstrate that \tNAME achieves 0.998 detection accuracy and outperforms state-of-the-art node-wise HT detection methods.
\tNAME extracts HT features without heuristic feature engineering.

%% file: tex/1_introduction.tex
\section{Introduction}
\label{sec:1_intro}

\IEEEPARstart{T}{he} demand for high-performance, low-cost, and power-saving ICs has been increasing, which makes supply chain protection a serious concern in the reality of the fourth industrial revolution.
To meet demand, the IC design process must be correspondingly secure.
Primary vendors often use third-party intellectual properties~(3PIP) and outsource parts of their products to third-party hardware design houses.
Utilizing 3PIP and outsourcing to the third-party vendors lead to the globalization and complexity of the supply chain, associated with the risk of unintended third parties' participation.

A \textit{hardware Trojan~(HT)} is emphasized as a threat in the supply chain~\cite{10.1145/2906147}.
An HT consists of two core components: trigger and payload and is often implemented as minute hardware with its trigger deactivated to evade inspections.
With the trigger deactivated and thus leaving its payload disabled, it acts as an HT-free IC.
When the HT's trigger is eventually activated, it may leak confidential information, tamper with functionality, and suspend devices.
%
HT detection at the design phase has been widely researched~\cite{9301614}.
In particular, gate-level netlists (hereinafter referred to as \textit{netlists}) are focused.
A structural feature-based HT detection method was proposed to show optimal performance \cite{7092434}, its merit being that it requires no simulation.
It also realizes the comprehensive and fine analysis of the target IC design.
The methods \cite{8952724,9424362} realize node-wise HT detection in netlists using machine learning~(ML) and show high detection performance.
However, feature-based ML methods have limitations in discovering an appropriate set of features.
Previous studies have adopted heuristic approaches to find structural features for HT detection.
The selected features are valid for known HTs, but skilled attackers can evade them.
It is a tremendous task to put upon structural feature-based HT detection to continuously extract effective HT features from the IC design when a new HT is found.
Thus, simply employing a structural feature-based approach is unfeasible for real world circuits.

To overcome these limitations of structural feature-based HT detection, a graph learning~(GL) method is introduced.
A circuit can be represented as a graph, such as Boolean networks~\cite{10.5555/548731}.
Likewise, a netlist is represented as a graph structure.
Its node shows an element of a circuit and its edge, a wire.
Graph structure is becoming an active research area in recent ML.
It is expected that GL extracts generalized features from netlists, an impossibility via manual feature engineering.
Considering that HTs are becoming more technical and sophisticated, GL is a promising approach.
GL-based HT detection \cite{9474174,9530566,muralidhar2021contrastive} distinguishes between normal circuits and HTs effectively.
The impracticalities of the existing methods consist of problem settings, such as circuit-wise classification and trigger-focused detection, and the fact that the features GL grasps is unknown.

In this paper, we propose \tNAME, a novel node-wise HT detection method based on GL.
The contributions of this paper can be summarized as follows:
\begin{itemize}
     \item The practical settings for HT detection are clarified by providing realistic scenarios in the hardware supply chain based on preliminary experiences of HT detection in netlists (Section~\ref{sec:3_motivation}).
     \item This paper bridges the gap between the known structural features of HTs and the representation capability of GNNs by clarifying what features a GNN model captures for HT detection (Section~\ref{sec:4_ht_features}).
     \item \tNAME, the hardware Trojan detection method in netlist using GL is proposed (Section~\ref{sec:5_proposed}), the theoretical background of which is described in Section~\ref{sec:4_ht_features}.
     \item \tNAME is evaluated through experiments. The experimental results demonstrate that \tNAME outperforms state-of-the-art HT detection methods.
     Additionally, by comparing a GNN-based HT detection method with simple node features, it is shown that the GNN model effectively extracts the features characterizing HTs from a given training dataset (Section~\ref{sec:6_evaluation}).
\end{itemize}

%% file: tex/2_backgrounds.tex
\section{Related Works}
\label{sec:2_backgrounds}



\begin{figure}[t]
    \center
    \includegraphics[width=1.00\linewidth]{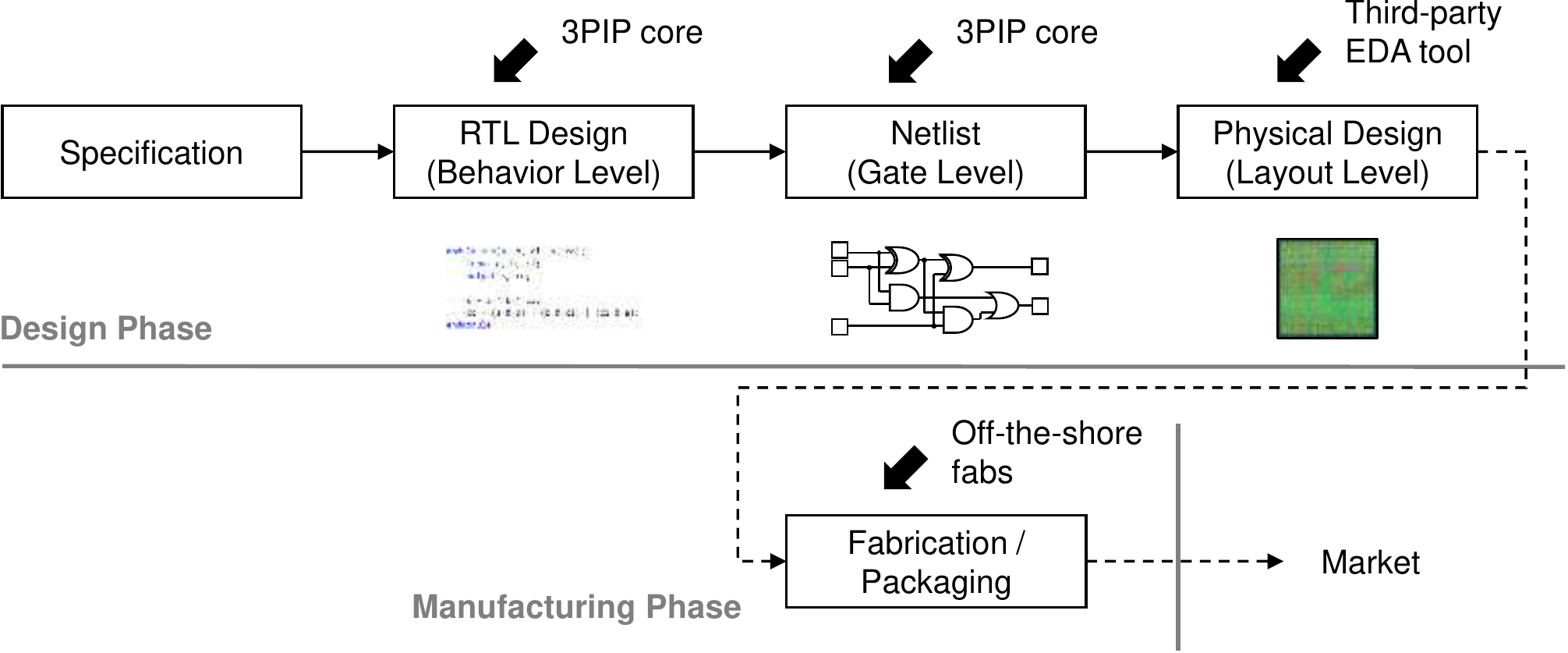}
    \caption{Typical IC design process. Malicious 3PIP cores as well as third-party EDA tools might be involved in the process.}
    \label{fig:supply-chain}
\end{figure}

\noindent
\textbf{HT detection}.
HT detection methods were reviewed in~\cite{10.1145/2906147}.
The typical IC design process is illustrated in Fig.~\ref{fig:supply-chain}.
In the design phase, specification is sequentially broken down into behavior level, gate level, and layout level.
3PIP core and third-party EDA tools have the opportunity to participate in the design phase.
Henceforth, malicious attackers may take advantage.
IC design is written in hardware description language~(HDL) and stored in an electrical design interchange format~(EDIF).
Skillful attackers who know the language and format can hide HTs to contaminate the IC design or modify the design information.
On the other hand, the attack in the manufacturing phase is difficult since the manufacturing system is working in real time and being controlled by a vendor-specific management process.
Therefore, attacking the design phase is a more realistic scenario than attacking the manufacturing phase.
This paper focuses on HTs inserted in the design phase, particularly in netlists rather than the more abstract level design such as register-transfer level (RTL), as the IC design described in RTL is ultimately translated into a netlist.

\noindent
\textbf{HT detection by logic testing}.
HTs are likely to be activated to evade from logic testing.
Focusing on rare activation conditions, early studies have proposed analytical detection methods based on truth tables or simulations \cite{10.1145/2508859.2516654,7086012}.
The weakness of hardware verification by logic testing is that it is time consuming to apply for large scale circuits, and that there exists a technique to make HTs stealthy to logic testing as proposed by DeTrust~\cite{10.1145/2660267.2660289}.

\noindent
\textbf{Feature-based HT detection}.
Another approach is the feature-based method proposed in \cite{7092434}, which requires no simulation and realizes comprehensive analysis with less time compared to HT detection by logic testing.
The method successfully detects HTs with a high accuracy rate using the structural features manually extracted from the Trust-HUB benchmark netlists~\cite{shakya2017benchmarking,Salmani2013}, which suggests that structural features exist which would specifically distinguish HTs from normal circuits.
Following the suggestion, several ML-based approaches have been proposed~\cite{8952724,9424362}.
According to \cite{9424362}, one of the first ML-based HT detection methods was proposed in \cite{conf/iolts/HasegawaOYT16}, which achieved high recall but insufficient accuracy by employing SVM and ANN to learn HT features.
Since then, new features and models have been actively investigated~\cite{conf/iscas/HasegawaYT17,7577739,9338932}.
In the past five years, HT detection methods using ML have achieved 90\% or more accuracy.
The weakness of the feature-based method is that it requires continuous feature updating.
It has been reported that a powerful feature-based method, COTD detection method~[26], would not be adequate in a certain application~\cite{conf/ismvl/ItoUH21}.
It suggests that HT-specific features should be searched when a new HT is discovered.
It is necessary to automate feature discovery process to real world problem.

\noindent
\textbf{GL-based HT detection}.
Graph learning is now becoming an active research area in ML~\cite{9039675}.
Since a netlist can be represented as a graph structure by translating an element of a circuit into a node and a wire into an edge, GL-based HT detection is a promising approach to break through the impasse of endless feature engineering.
Ref.~\cite{muralidhar2021contrastive} effectively detects HTs from netlists using GL.
GL-based HT detection no longer requires feature engineering.
Features are automatically extracted in the learning process.
As reported in \cite{9474174}, simple GL-based HT detection methods bear a weakness that they do not point out where HTs exist, but only identify whether the design is compromised or not.
For practical use, detected HT should be presented with evidence so that the user can trust the detection result. 
The only solution to this problem would be explaining the GL model to those who are not well-versed in HTs. 
In \cite{muralidhar2021contrastive}, only the trigger part is detected, leaving its critical payload part unnoticed.
Also, their proposed INF contains several normal gates, which may hurt generalization performance.
Therefore, different embedding approaches, such as optimizing node features based on the representation capability of the GL model, should be considered.


\section{Preliminaries}
\label{subsec:2_2_graph_learning}


\noindent
\textbf{General model}.
The general model of \tNAME is the Graph Neural Network~(GNN).
The concept of GNN is introduced in \cite{1555942} and is now employed in a variety of fields~\cite{9039675}.

The MPNN model~\cite{10.5555/3305381.3305512} has proposed a unified framework for graph convolution operation, and many models can be considered with this framework.
Let $G = (V, E)$ be a graph with a set $V$ of nodes and a set $E$ of edges.
A feature vector $\vecx_v$ is assigned for $v \in V$, which is referred to as an initial feature vector that represents the property of a node $v$.
Also, a label $y_v \in \mathbb{R}$ represents the class of the node $v$.
In general, GNN employs the neighborhood aggregation~(a.k.a. \textit{message passing}) mechanism, in which node feature vectors are exchanged between nodes, and updates (or {combines}) them using an updating function.
This mechanism is expressed as follows~\cite{conf/iclr/XuHLJ19}:
\begin{align}
    \vecm^{(l)}_{v} & = \mathsf{AGGREGATE}^{(l)} \left( \left\{ \vechl_{u}: \forall u \in \mathcal{N}(v) \right\} \right) \label{eq:gnn-agg} \\
    \vech^{(l+1)}_{v} & = \mathsf{COMBINE}^{(l)} \left( \vechl_{v}, \vecm^{(l)}_{v} \right), \label{eq:gnn-comb}
\end{align}
where $\mathcal{N}(v)$ represents a set of nodes adjacent to a node $v$, and $\mathsf{AGGREGATE}^{(l)}(\cdot)$ (resp. $\mathsf{COMBINE}^{(l)}(\cdot)$) are the message function (resp. the update function) at the $l$-th layer.
Such a layer is referred to as a \textit{GNN layer} in this paper.
$\vechl_{v}$ is a hidden feature vector of $v$ initialized by the initial feature vector as $\vech^{(0)}_{v} = \vecx_v$, and $\vecm^{(l)}_{v}$ denotes the message exchanged between nodes.
In (\ref{eq:gnn-agg}), the message function $\mathsf{AGGREGATE}^{(l)}(\cdot)$ aggregates the feature vectors of the nodes adjacent to $v$ and generates the message vector $\vecm^{(l)}_{v}$.
Then, in (\ref{eq:gnn-comb}), the update function $\mathsf{COMBINE}^{(l)}(\cdot)$, which is a learnable function, combines the node feature vector $\vechl_{v}$ and the message vector $\vecm^{(l+1)}_{v}$.
Finally, the node feature vector $\vechl_{v}$ for each node at the $l$-th GNN layer is updated.
The functions $\mathsf{AGGREGATE}^{(l)}(\cdot)$ and $\mathsf{COMBINE}^{(l)}(\cdot)$ are different from model to model, which results in the capability of representation of a node.

Let $\vecz_{v} = \vechL_{v}$ be a final output for the node feature vector of $v$, where $L$ is the number of GNN layers.
In task-specific processing, the ML model for graph can be divided into a graph encoder part and prediction part.
Let $\funcfg$ and $\funcfc$ be a graph encoder model and prediction model, respectively.
The graph encoder model can be expressed as $\funcfg(\vecx_v) = \vecz_v$ that implies (\ref{eq:gnn-agg}) and (\ref{eq:gnn-comb}).
The prediction model can be expressed as $\funcfc(\vecz_v) = y_v$.
For example, the optimization problem for binary classification is expressed as follows:
\begin{equation}
    \min \mathcal{L} \left( y_v, \funcfc( \vecz_{v} ) \right),
\end{equation}
where $\mathcal{L}$ is a loss function, and $\funcfc$ is a classification model, such as a fully-connected neural network.

%% file: tex/3_motivation.tex
\section{Motivations}
\label{sec:3_motivation}


\smallskip
\noindent
\textbf{Threat model}.
As illustrated in Fig.~\ref{fig:supply-chain}, there are many opportunities for attackers to be involved in the hardware supply chain, such as providing malicious 3PIP cores or invading as untrusted off-the-shore fabs.
In particular, there are more opportunities for attackers to insert HTs in the design phase as addressed by our threat model.
Specifically, attackers may insert HTs into the 3PIP cores and provide them to the primary vendor.
Attackers may also invade the design house and directly insert HTs with malicious intent.
This scenario housed in the recent hardware supply chain involves many employees, partners, 3PIPs, and off-the-shore design houses.
Henceforth, the supply chain becomes unsecure, providing loopholes for malicious attackers to gain entry.

\smallskip
\noindent
\textbf{Our motivations}.
There are three motivations in this paper, and this section clarifies them from the following perspectives:
\begin{itemize}
 \item \tMotivationA describes why we target netlists and the \tNodeLevel HT detection (Section~\ref{subsec:3_1_mot1}).
 \item \tMotivationB describes why we employ graph learning for HT detection (Section~\ref{subsec:3_2_mot2}).
 \item \tMotivationC describes why we introduce domain knowledge for graph learning (Section~\ref{subsec:3_3_mot3}).
\end{itemize}

\subsection{\tMotivationA: Gate-Level Netlists and \TTNodeLevel HT Detection}
\label{subsec:3_1_mot1}


\noindent
\textbf{HT detection in netlists.}
As illustrated in Fig.~\ref{fig:supply-chain}, the design phase is roughly broken down into four levels: specification, behavior level, gate level, and layout level.
%
Any IC design must pass through the \textit{gate level}.
Even if some 3PIPs are provided as the \textit{behavior-level} description, they are synthesized to the \textit{gate-level} description.
This can also be done of the \textit{layout level}.
However, it includes location information, which is not needed for behavior analysis.
%
Netlists at the gate level are common and useful in the IC design phase.
In fact, an HT detection service commercially available is detailed in \cite{htfinder}, which targets netlists.
Therefore, this paper focuses on HT detection in netlists.

\noindent
\textbf{\TNodeLevel detection}.
There are two approaches in HT detection in netlists: \tCircuitLevel and \tNodeLevel detection.
The \tCircuitLevel detection identifies whether the IC design includes an HT or not.
Alternatively, \tNodeLevel detection identifies which node is a part of an HT.
If a \tCircuitLevel HT detection system finds that an HT may exist in the IC design, the user may doubt such an alert.
The user must determine whether or not the product should be re-designed based on the result of the detection system.
Also, the cause of the HT insertion should be carefully analyzed if an HT actually exists.
\TNodeLevel detection solves the problem.
The results show which node could be a part of an HT.
If a \tNodeLevel detection method achieves an acceptable rate in detection accuracy, its result must support further analysis and decision-making.
Therefore, \tNodeLevel detection is more helpful for practical use.

\subsection{\tMotivationB: Graph Learning}
\label{subsec:3_2_mot2}

As mentioned in Section~\ref{subsec:2_2_graph_learning}, GNNs are expected to capture the generalized features of graph-structured data.
Since many engineers are now joining the hardware design community because of the spread of open-source projects such as RISC-V~\cite{risc-v}, HT detection that does not require special knowledge of HTs is needed.
Additionally, HT structure can be easily transformed.
For example, the transduction method~\cite{35836} transforms a logic design into a different structure while keeping the original functionality.
If we can represent a circuit with a generalized form, the circuits with the same functionality would be embedded into similar latent spaces; effect of the automatic feature extraction is significant.

To study the features of HTs, we can collect many HTs by using automatic HT generation tools~\cite{conf/date/CruzHMB18,conf/isvlsi/YuLO19} proposed very recently.
However, even if many types of HTs could be collected, it is too difficult to extract a set of comprehensive HT features.
GL-based HT detection is expected to extract the features included in the training dataset automatically.

\subsection{\tMotivationC: Domain Knowledge}
\label{subsec:3_3_mot3}

To make the HT detection system reliable, evidence and theoretical background information should be made clear.
Although feature engineering may not be required in GL, knowledge of HT features is vital.
Existing studies have demonstrated that their proposed features are useful for HT detection, and work effectively for a certain set of HTs.
Therefore, knowledge of HTs still comes of aid for GL-based HT detection.

Introducing the domain knowledge is different from automatic feature extraction.
The automatic feature extraction, which can be performed by GL, is a process to choose a set of efficient features from a set of initial features of nodes in a given training dataset.
If sufficient initial features are not provided to the initial feature vector of a graph, the performance of GL may saturate at an insufficient level.
The role of domain knowledge is the selection of a set of sufficient initial features that represent the nodes in a netlist.
By selecting common features that can be observed in a wide variety of HTs as initial features based on domain knowledge, feature extraction by GL is expected to work effectively.

Optimizing initial feature vectors based on domain knowledge is helpful.
As described in Section~\ref{subsec:2_2_graph_learning}, the hidden feature vector for a node $v$ initialized as $\vech^{(0)}_v = \vecx_v$.
Even though GL can automatically learn the representation of nodes in a graph, the node features given as an initial feature vector significantly affect the performance of the subsequent task.
In \cite{duong2019on}, it has been demonstrated through experiments that GNNs work well if there is a strong correlation between node feature vectors and node labels.
Thus, the node features are introduced to sufficiently represent known HT features and HT-related circuit features.
%
This paper aims to find efficient features representing the characteristics of a node for GL-based HT detection.



%% file: tex/4_ht_features.tex
\section{HT Features}
\label{sec:4_ht_features}


\subsection{HT Features at Gate-Level Netlists}
\label{subesc:4_1_HT_features}

In this subsection, we refer to several structural feature-based HT detection methods and categorize what they learn from the perspective of graph-structured data.
According to \cite{9301614}, there are several approaches for HT detection in netlists using a variety of features.
Hereafter, we refer to the three references~\cite{conf/iolts/KuriharaT21}, \cite{8624727}, and \cite{8702462} to introduce the representative features based on the studies above.

In \cite{conf/iolts/KuriharaT21}, 36 structural features are employed for HT detection in total.
Some of the features are introduced at one of the earliest studies~\cite{conf/iolts/HasegawaOYT16,conf/iscas/HasegawaYT17} that are inspired by \cite{7092434}.
They extract feature values from each net in a netlist from the viewpoint of fan-ins, neighbor circuit elements, and the minimum distance to the specific circuit elements.
Ref.~\cite{conf/iolts/KuriharaT21} further introduces the pyramidal structure-based features called \textit{fan\_in\_uxdy}, and they improve detection performance.
Table~\ref{tbl:ht-features-kurihara} shows the 36 features presented in \cite{conf/iolts/KuriharaT21}.
The `category' column is introduced later.
The features mainly focus on the structural features of a netlist, that is, the topological features of a graph.

\begin{table}[t]
    \centering
    \caption{HT features employed in \cite{conf/iolts/KuriharaT21}.}
    \label{tbl:ht-features-kurihara}
    \scalebox{0.92}{
    \begin{tabular}{c|p{40mm}|l} \hline
        No. & Feature & Category \\ \hline\hline
        1--2 & No. of fan-ins up to 4 and 5-level away from the input side & \circled{1}~Degree \\ 
        3 & No. of flip-flops up to 4-level away from the input side & \circled{2}~Neighboring node \\ 
        4--5 & No. of flip-flops up to 3 and 4-level away from the output side & \circled{2}~Neighboring node \\ 
        6--7 & No. of loops up to 4 and 5-level on the input side & \circled{3}~Relative position \\ 
        8 & Min. level to any primary input & \circled{3}~Relative position \\ 
        9 & Min. level to any primary output & \circled{3}~Relative position \\ 
        10 & Min. level to any flip-flops from the output side & \circled{3}~Relative position \\ 
        11 & Min. level to any multiplexer from the output side & \circled{3}~Relative position \\ 
        12--36 & Pyramidal structure-based feature within 4 levels & \circled{4}~Surrounding structure \\ \hline
    \end{tabular}
    }
\end{table}

In \cite{8624727}, 15 features are employed, which is shown in Table~\ref{tbl:ht-features-hoque}.
Different from \cite{conf/iolts/KuriharaT21}, the features mainly focus on the functional behavior such as static probability and signal rate (No.~6--15 in Table~\ref{tbl:ht-features-hoque}).
The functional behavior-based features reflect the functionality of a netlist, such as rare or frequent transition of a signal.
Although the structural features can capture the structure of a set of nodes, it does not explicitly capture the behavior of the circuit.
Functional behavior-based features solve the problem, and they capture the behavior of the circuit explicitly.

\begin{table}[t]
    \centering
    \caption{HT features employed in \cite{8624727}.}
    \label{tbl:ht-features-hoque}
    \scalebox{0.92}{
    \begin{tabular}{c|p{40mm}|l} \hline
        No. & Feature & Category \\ \hline\hline
        1--2 & No. of immediate fan-in and fan-out& \circled{1}~Degrees \\ 
        3 & Cell type driving the net & \circled{2}~Neighboring node \\ 
        4--5 & Min. distances from PI and PO & \circled{3}~Relative position \\ 
        6 & Static probability & \circled{5}~Functional behavior \\ 
        7 & Signal rate & \circled{5}~Functional behavior \\ 
        8 & Toggle rate & \circled{5}~Functional behavior \\ 
        9 & Min. toggle rate of the fan-outs & \circled{5}~Functional behavior \\ 
        10 & Entropy of the driver function & \circled{5}~Functional behavior \\ 
        11--15 & Lowest, highest, average, std. and dev. of controllability & \circled{5}~Functional behavior \\ \hline
    \end{tabular}
    }
\end{table}

In \cite{8702462}, six testability metrics-based features are employed.
Table~\ref{tbl:ht-features-kok} shows the six features presented in \cite{8702462}.
These features are known as \textit{SCOAP} values and utilized inherently for evaluating the testability of a circuit~\cite{1585245}.
The recent study~\cite{7577739} has first introduced them to HT detection, and they are often employed for HT detection.
Since an HT is hard to observe and easy to control from outside the circuit, the \textit{SCOAP} values are reasonable for HT detection.

\begin{table}[t]
    \centering
    \caption{HT features employed in \cite{8702462}.}
    \label{tbl:ht-features-kok}
    \scalebox{0.92}{
    \begin{tabular}{c|p{40mm}|l} \hline
        No. & Feature & Category \\ \hline\hline
        1--2 & Controllability (CC0 and CC1) & \circled{5}~Functional behavior \\ 
        3 & Observability (CO) & \circled{5}~Functional behavior \\ 
        4--5 & Sequential controllability (SC0 and SC1) & \circled{5}~Functional behavior \\ 
        6 & Sequential observability (SO) & \circled{5}~Functional behavior \\ \hline
    \end{tabular}
    }
\end{table}

\smallskip
\noindent
\textbf{Toward GL}.
Based on the observation of several existing HT detection methods, we analyze the features and categorize them from the viewpoint of GL. 

\begin{itemize}
    \item \circled{1}~\textbf{Degree} shows the degree of a node in a graph, which is related to the number of edges connected to the node.
    \item \circled{2}~\textbf{Neighboring node} shows the types of nodes neighboring a target node.
    \item \circled{3}~\textbf{Relative position} shows the relative position to the specific sets of nodes.
    \item \circled{4}~\textbf{Surrounding structure} shows the surrounding structure from a target node.
    \item \circled{5}~\textbf{Functional behavior} shows the \emph{functional behavior} of a node i.e. characterizes the behavior of a set of nodes.
    An example is static probability.
    Typically, the trigger circuit of an HT rarely activates the trigger signal.
    The node outputting trigger signal is rarely activated.
    Such behavior cannot be observed by simply analyzing structural features.
\end{itemize}


The following section analyzes the representation capability of graph encoding models from the perspective of the categories \circled{1}--\circled{5}.

\subsection{Representing HT Features with GL}

Here we bridge the gap between the known HT features and GL.
As described in the previous section, the HT features are classified into five categories.
We demonstrate how GL captures these features in a formal manner.

\smallskip
\noindent
\circled{1}~\textbf{Degree}:
Fan-ins and fan-outs of a node are useful information for HT detection.
For example, a trigger circuit composed of a combinatorial circuit often has many fan-ins at two or three levels from the trigger signal wire to implement rare conditions.
%
%
The node degrees can be directly assigned to the initial feature vector to clearly make the graph encoder model identify node degrees.

\smallskip
\noindent
\circled{2}~\textbf{Neighboring node}:
The information about the types of nodes neighboring a target node is helpful for HT detection.
To utilize the information by GL, a node type is assigned to the initial feature vector.
%
%

The node types are the most fundamental features for representing the characteristics of nodes in a netlist, as used in \cite{9530566} and \cite{muralidhar2021contrastive}.
In this paper, the node types are used as the features of a baseline, the details of which is presented later.

\smallskip
\noindent
\circled{3}~\textbf{Relative position}:
When the surrounding structures of two subgraphs are identical, the graph encoder model that considers only the feature categories \circled{1} and/or \circled{2} cannot identify the subgraphs.
According to (\ref{eq:gnn-comb}), the final node feature vector of $v$ is expressed as $\vecz_v = \vech^{(L)} = \mathsf{COMBINE}^{(L)} \left( \vech^{(L-1)}_{v}, \vecm^{(L-1)}_{v} \right)$.
When $\vech^{(L-1)}_{v} = \vech^{(L-1)}_{u}$ and $\vecm^{(L-1)}_{v} = \vecm^{(L-1)}_{u}$ for two nodes $v$ and $u$, the node feature vectors of the two nodes are identical, i.e., $\vecz_v = \vecz_u$.
In a netlist, such a case can appears in a ring oscillator that is composed of multiple inverter gates forming a loop structure or a multi-bit counter that is composed of several 1-bit counters.
These circuits can be used as a payload circuit or a sequential trigger circuit of an HT.

To overcome the limitation of GNN, a position-aware GNN model has been proposed in \cite{conf/icml/YouYL19}.
In the method, a random subset of $k$ nodes are chosen as \textit{anchor-sets} and their node feature vectors are included in the target node feature vector.
Inspired by \cite{conf/icml/YouYL19}, we state the following proposition:

\begin{proposition}
    \label{prop:f3:gnn_loops}
    Let $G_{\mathsf{anchor}}$ be a graph where an initial feature vector $\vecx_{v}$ that includes information on \textit{anchor-sets} is assigned to each node $v$.
    Let $\phi \colon \mathbb{R}^d \rightarrow \mathbb{R}^d$ be a GNN layer, and let $\funcFg$ denote a set of GNN models that consist of more than one layer $\phi$ and are injective.
    There exists a graph encoder model $\funcfg \in \funcFg$ that takes $G_{\mathsf{anchor}}$ as an input and identifies the difference of two nodes in $G_{\mathsf{anchor}}$ in terms of position in a graph.
\end{proposition}


The proof of Proposition~\ref{prop:f3:gnn_loops} is shown in Appendix~\ref{apdx:prf:prop:f3:gnn_loops}.
According to \cite{conf/icml/YouYL19}, choosing a set of anchor nodes is a difficult problem for an inductive learning task.
From the viewpoint of netlists, primary inputs and outputs are reasonable candidates for \textit{anchor-sets} making them useful for HT detection.
Therefore, a graph encoder model can best identify the positional feature of a node by employing minimum distances to any primary inputs and outputs as initial feature values.

\begin{table}[t]
    \centering
    \caption{GL-based methods for netlists.}
    \label{tbl:graph-modeling-methods}
    \begin{tabular}{l|l} \hline
        Graph & Methods \\ \hline\hline
        Undirected graph & GNN-RE~\cite{9530566}, GATE-Net~\cite{muralidhar2021contrastive} \\ 
        Directed graph & \cite{9424256}, GNN4TJ~\cite{9474174} \\ \hline
    \end{tabular}
\end{table}

\smallskip
\noindent
\circled{4}~\textbf{Surrounding structure}:
How a netlist is represented as a graph structure is one of the most problematic issues in GL.
There are two forms of representing a netlist: undirected graph and directed graph.
The edge between two nodes in a graph is directional in a directed graph, whereas it is not in an undirected graph.
The signal wires in a netlist are inherently directional, and thus directed graph seems to be suitable to represent a netlist.
However, a directed graph cannot sufficiently represent a netlist for graph encoder models.
With a directed graph, the aggregation function gathers only one direction of each edge, and thus the nodes connected to the opposite side are ignored.
Table~\ref{tbl:graph-modeling-methods} shows the graph models employed in recent processing methods using GNNs.
According to \cite{9530566}, which aims to represent netlists for multiple downstream tasks, representing a netlist as an undirected graph is more efficient than as a directed graph.
It should be noted that GNN4TJ~\cite{9474174}, which employs directed graphs, mainly focuses on HTs written in RTL.
Therefore, a directed graph may be inefficient for netlists.

\begin{figure}[t]
    \center
    \includegraphics[width=0.80\linewidth]{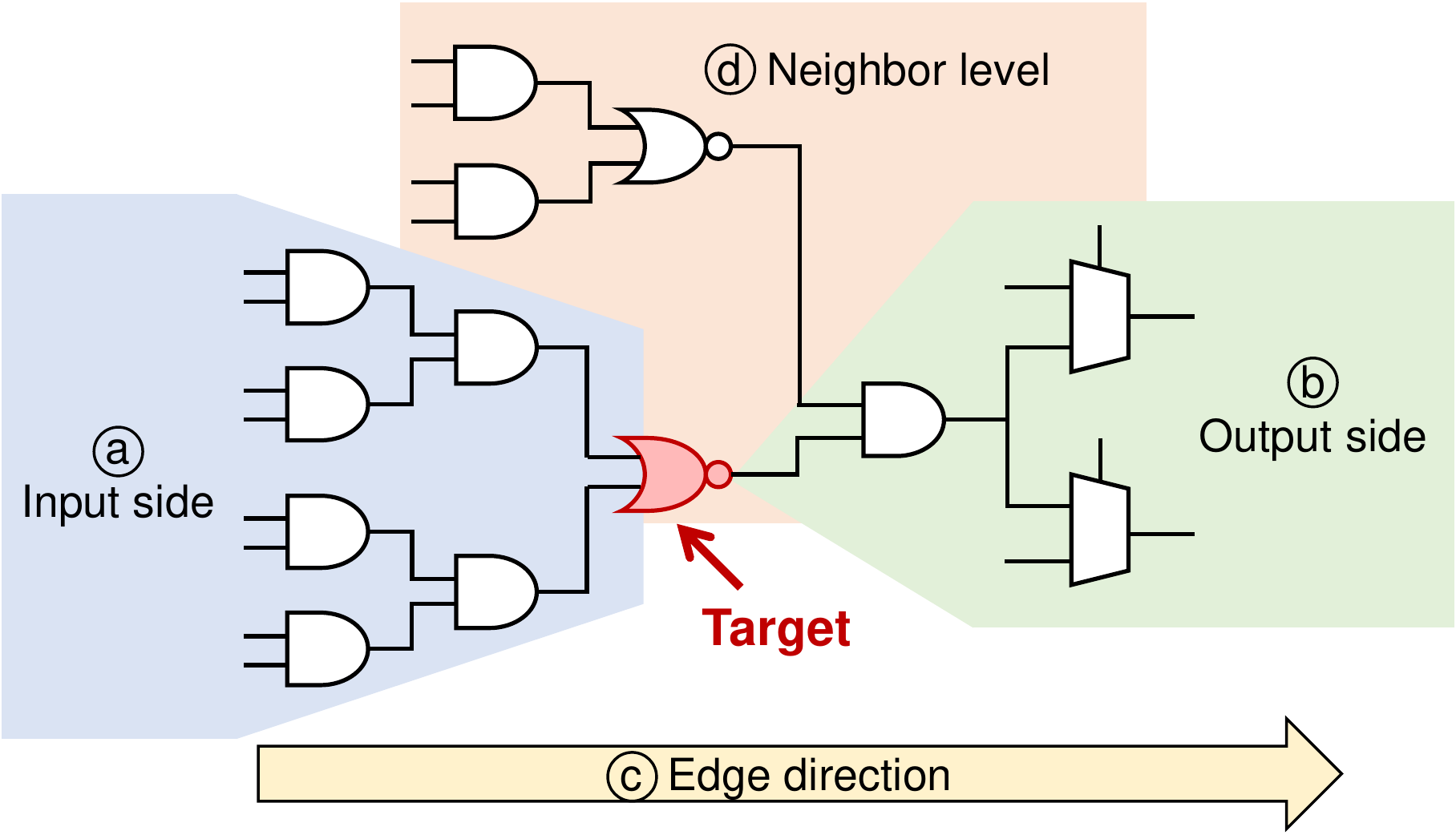}
    \caption{Example of graph-related features for a netlist.}
    \label{fig:graph-rep-model}
\end{figure}

The representation capability of the directed and undirected graphs is broken down, and a new model for representing a netlist is proposed.
Four points of the graph-related features for a netlist are as follows:
\begin{itemize}
    \item \circleedged{a}~\textbf{Input side}: The structure of the input side from target node $v$.
    \item \circleedged{b}~\textbf{Output side}: The structure of the output side from target node $v$.
    \item \circleedged{c}~\textbf{Edge direction}: The direction of an edge. Namely, the direction of a wire in a netlist.
    \item \circleedged{d}~\textbf{Neighbor level}: The structure of the neighbor level from target node $v$.
\end{itemize}
Fig.~\ref{fig:graph-rep-model} illustrates an example of the graph-related features using a circuit.
In this figure, we focus on the `target' node colored in red.
\circleedged{a}~\textbf{Input side} corresponds to the area shaded in light blue.
The nodes in this area are connected to the input side of the target gate.
\circleedged{b}~\textbf{Output side} corresponds to the area shaded in light green.
The nodes in this area are connected to the output side of the target gate.
\circleedged{c}~\textbf{Edge direction} shows whether the edge direction is preserved or not in the modeled graph.
\circleedged{d}~\textbf{Neighbor level} corresponds to the area shaded in orange.
The neighbor level area can be reached by traversing some hops toward the output side and then some hops backward the input side or vice-versa.

\begin{table}[t]
    \centering
    \caption{Graph structure and its representation capability in a graph encoder model.}
    \label{tbl:graph-structure}
    \scalebox{0.8}{
    \begin{tabular}{l|cccc} \hline
                                    & \circleedged{a} & \circleedged{b} & \circleedged{c} & \circleedged{d} \\
        Graph type                  & Input & Output & Edge & Neighbor \\
                                    & side & side & direction & level \\ \hline\hline
        {Directed}   & \checkmark &  & \checkmark &  \\
        {Undirected} & \checkmark & \checkmark &  & \checkmark \\ \hline
        Undirected                  & \mulrowchk & \mulrowchk & \mulrowchk & \mulrowchk \\
        {\scriptsize ~(with directional edge attribute)} & & & & \\ \hline
    \end{tabular}
    }
\end{table}

Next, we summarize the representation capability of graph models.
Table~\ref{tbl:graph-structure} shows the graph structures and their representation capability in a graph encoder model.
A directed graph can represent the input side and the edge direction of a netlist when a graph encoder model aggregates the nodes of the input side.
However, the nodes of the output side are not aggregated in this situation.
Similarly, the neighbor levels are not considered unless there is a loop structure.
On the other hand, an undirected graph can represent both the input and output sides.
Due to the both-side aggregation, it can also represent the nodes in neighbor levels.
However, it lacks directional information.

Then, we propose the directional edge attribute to an undirected graph, called \textit{edge-attributed undirected graph~(EAUG)}.

\begin{definition}[Edge-attributed undirected graph]
    An edge-attributed undirected graph (EAUG) is a graph $G_{\mathsf{EAUG}} = (V, E, X, D)$ that represents a netlist with a set of initial node vectors $X = \{ \vecx_{v}: v \in V \}$ and a set of edge attribute vectors $D = \{ \vecd_{u \rightarrow v}: u, v \in V \}$.
\end{definition}

The EAUG overcomes the shortcoming of the undirected graph.
Let $e_{u \rightarrow v}$ denote an edge from a node $u$ to another node $v$.
$\vecd_{u \rightarrow v} \in \mathbb{R}^d$ is an edge attribute vector assigned to $e_{u \rightarrow v}$.
In a typical undirected graph, the edges $e_{u \rightarrow v}$ and $e_{v \rightarrow u}$ are not distinguished.
Furthermore, when feature vectors are assigned to the edges of an undirected graph, $\vecd_{u \rightarrow v}$ and $\vecd_{v \rightarrow u}$ are usually the same.
In the EAUG, we distinguish two edges $e_{u \rightarrow v}$ and $e_{v \rightarrow u}$ by assigning different edge attributes as $\vecd_{u \rightarrow v} \ne \vecd_{v \rightarrow u}$
Then, we can state the following proposition.

\begin{proposition}
    \label{prop:f5:gnn_eaug}
    There exists a graph encoder model $\funcfg \in \funcFg$ that takes an edge-attributed undirected graph $G_{\mathsf{EAUG}}$ as an input and represents all the graph-related features \circleedged{a}--\circleedged{d}.
\end{proposition}

The proof of Proposition~\ref{prop:f5:gnn_eaug} is shown in Appendix~\ref{apdx:prf:prop:f5:gnn_eaug}.
For example, we assign a vector $(1, 0)$ to a forward direction edge and $(0, 1)$ to a backward direction edge of a netlist.
Then, we can construct an EAUG for a given netlist.




  




\smallskip
\noindent
\circled{5}~\textbf{Functional behavior}:
In this section, we observe the global behavior of a circuit.
At first, we consider the behavior of each node.
We call a function assigned to each node \emph{node behavior}, which is identified by only the node.
However, we can only observe the local behavior of each node by simply considering the \emph{node behavior}.
To observe the global behavior of a circuit, we need a new feature that takes into account the \emph{node behavior} of the neighboring nodes.
Then, we introduce the \emph{functional behavior}, which characterizes the behavior of a set of nodes.
The \emph{functional behavior} of a target node $v$ refers to a feature computed using an arbitrary function that takes node behaviors of $v$ and its neighboring nodes $\funcNk(v)$, where $\funcNk(\cdot)$ is a set of neighboring nodes within $k$ hops from a node.
In a netlist, there are several metrics that characterize the behavior of a set of nodes by assigning a function to each node, such as static probability and switching probability.
We consider such metrics by the \emph{functional behavior}.

\begin{figure}[t]
    \center
    \includegraphics[width=0.98\linewidth]{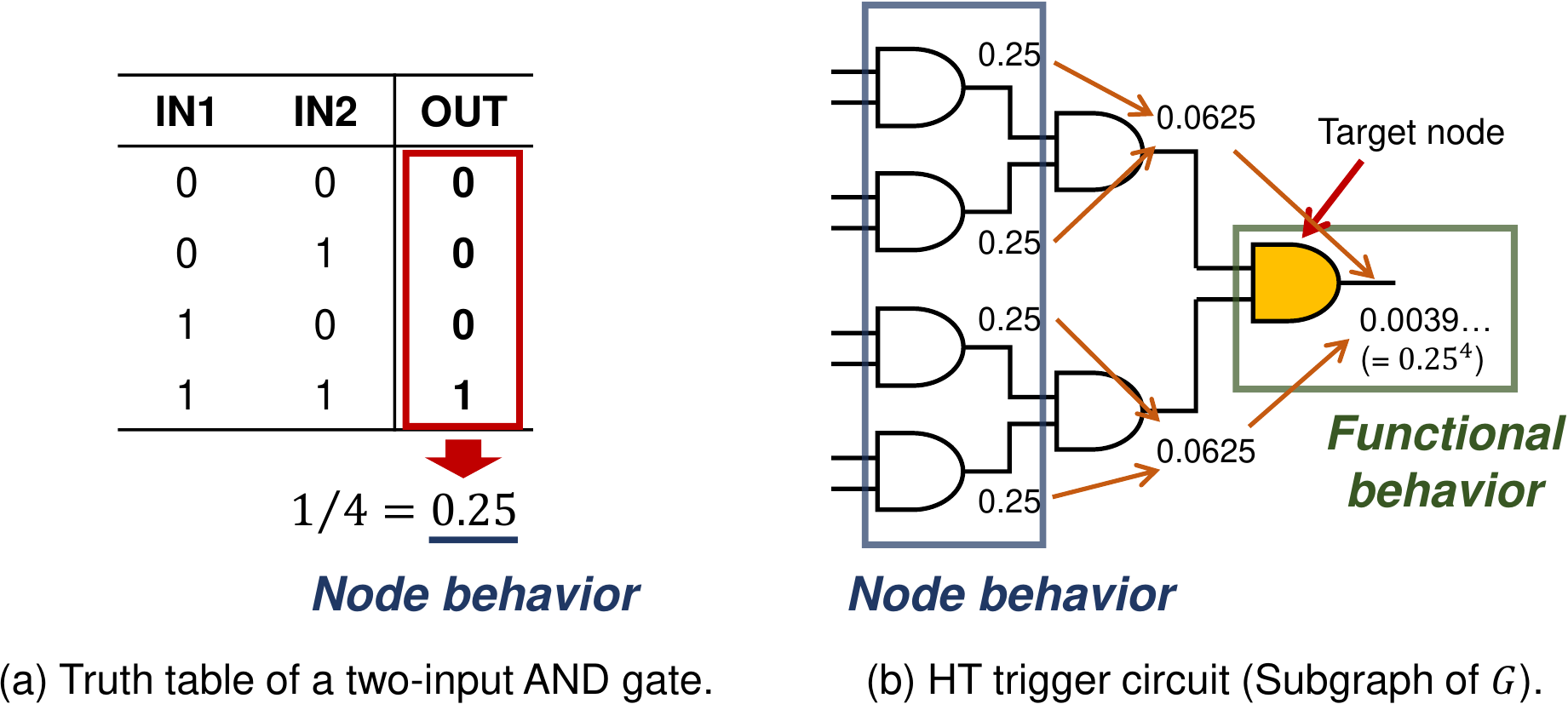}
    \caption{Example of a node behavior and a functional behavior.}
    \label{fig:example-of-functional-behavior}
\end{figure}

Specifically, static probability is an example of the \emph{functional behavior}, which shows the probability that a signal holds a logic value of 1 during a period.
Fig.~\ref{fig:example-of-functional-behavior} illustrates the example of an HT trigger circuit that consists of seven two-input AND gates.
First, we refer to the truth table to consider the local behavior of a two-input AND gate.
Here, we regard the probability of outputting 1 as the \emph{node behavior}.
In this case, the \emph{node behavior} of a two-input AND gate is 0.25, as shown in Fig.~\ref{fig:example-of-functional-behavior}(a).
Then, we consider the HT trigger circuit depicted in Fig.~\ref{fig:example-of-functional-behavior}(b).
Starting from the \emph{node behavior} values, we can calculate static probabilities of the target node.
We regard the static probability as the \emph{functional behavior}, which characterizes the functionality of a set of neighboring nodes within two hops from the target node.
As exemplified above, a \emph{functional behavior} also characterizes the behavior of HTs well, especially HT triggers as shown in Fig.~\ref{fig:example-of-functional-behavior}(b).

To formulate the \emph{functional behavior}, let $\funcType(v)$ be a node type and $\funcBehavior(v)$ be the \emph{node behavior} of node $v$.
The \emph{functional behavior}, $\funcFB(v)$ of a node $v$ can be calculated as follows:
\begin{equation}
    \label{eq:functional_behavior}
    \mathsf{fb}(v) \leftarrow  \Gamma_{\funcType(v)}\left( \funcBehavior(v), \{\funcBehavior(u): \forall u \in \funcN_{k}(v)\} \right),
\end{equation}
where $\Gamma_{\funcType(v)}$ is an arbitrary function parameterized by $\funcType(v)$.
The following proposition states that a graph encoder model that represents such a \emph{functional behavior} exists.

\begin{proposition}
    \label{prop:f6-3:gnn_functional_behavior}
    Let $G'$ be a graph constructed as a $\Geaug$ where the node type and a feature value related to a \emph{functional behavior} is assigned to each node $v$.
    There exists a graph encoder model $\funcfg \in \funcFg$ that takes $G'$ as an input and identifies the \emph{functional behavior} $k$-hop away from each node $v$ for a given $k$.
\end{proposition}

The proof of Proposition~\ref{prop:f6-3:gnn_functional_behavior} is given in Appendix~\ref{apdx:prf:prop:f6-3:gnn_functional_behavior}.

\smallskip

We consider the feature extraction model that extracts the graph structure, initial feature vectors, and edge attribute vectors for a given netlist.
Let $\Lambda$ be a feature extraction model that takes a netlist, which is denoted by $\vartheta$, and outputs a tuple $(V, E, X, D)$ of a set of nodes, set of edges, set of node feature vectors, and set of edge attribute vectors.
In some cases, $X$ or $D$ can be an empty set.
Let $\funcfgwith{\Lambda(\vartheta)} (v)$ be a graph encoder model with respect to $\Lambda(\vartheta) = (V, E, X, D)$ for obtaining the node feature vector of the node $v$ during the calculation.
$\funcfg(v)$ implicitly refers to $V$, $E$, $X$, and $D$ of a netlist $\vartheta$.
If $\exists u, v \in V, u \ne v, \funcfgwith{\Lambda(\vartheta)} (u) \ne \funcfgwith{\Lambda(\vartheta)} (v)$, it means that the combination of the models $\Lambda$ and $\funcfg$ can distinguish the two nodes $u$ and $v$.
$P_{\vartheta}(\Lambda, \funcfg)$ denotes the representation capability for the nodes in a given netlist $\vartheta$ when using $\Lambda$ as a feature extraction model and $\funcfg$ as a graph encoder model.
Here, the representation capability for a netlist is defined.

\begin{definition}[Representation capability for a netlist]
    Let $\Lambda_1$ and $\Lambda_2$ be feature extraction models, and let $\funcfg^{(1)}$ and $\funcfg^{(2)}$ denote graph encoder models.
    The representation capability $P_{\vartheta}(\Lambda_1, \funcfg^{(1)})$ is greater than $P_{\vartheta}(\Lambda_2, \funcfg^{(2)})$ if and only if $\forall u, v \in V, u \ne v, \funcfgwith{\Lambda_2(\vartheta)}^{(2)} (u) \ne \funcfgwith{\Lambda_2(\vartheta)}^{(2)} (v) \implies \funcfgwith{\Lambda_1(\vartheta)}^{(1)} (u) \ne \funcfgwith{\Lambda_1(\vartheta)}^{(1)} (v)$, and the relationship of the two representation capabilities is denoted as follows:
    \begin{equation}
        P_{(\vartheta)}(\Lambda_1, \funcfg^{(1)}) \succeq P_{(\vartheta)}(\Lambda_2, \funcfg^{(2)}).
    \end{equation}

    In particular, $P_{\vartheta}(\Lambda_1, \funcfg^{(1)})$ is \textbf{strictly} greater than $P_{\vartheta}(\Lambda_2, \funcfg^{(2)})$ if and only if $P_{(\vartheta)}(\Lambda_1, \funcfg^{(1)}) \succeq P_{(\vartheta)}(\Lambda_2, \funcfg^{(2)})$ and $\exists u, v \in V, u \ne v, \funcfgwith{\Lambda_1(\vartheta)}^{(1)} (u) \ne \funcfgwith{\Lambda_1(\vartheta)}^{(1)} (v)$ but $\funcfgwith{\Lambda_2(\vartheta)}^{(2)} (u) = \funcfgwith{\Lambda_2(\vartheta)}^{(2)} (v)$.
\end{definition}

For comparison of the representation capability of graph encoder models, we introduce two feature extraction models for a netlist: baseline and netlist feature extraction models.

\begin{definition}[Baseline feature extraction model]
    A baseline feature extraction model $\Lambda_{\mathsf{baseline}}(\vartheta) = (V, E, X, D)$ extracts a set of node initial feature vectors $X$ that indicate the node types of each node $v \in V$ and does not extracts any edge attribute, i.e., $D = \emptyset$.
\end{definition}

The model that extracts the features belong to the feature category \circled{2} is a baseline feature model.

\begin{definition}[Netlist feature extraction model]
    A netlist feature extraction model $\Lambda_{\mathsf{netlist}}(\vartheta) = (V, E, X, D)$ extracts a set $X$ of node initial feature vectors that indicate the node degrees, node types, the features of relative position to anchor-sets, and the features of functional behavior of each node $v \in V$ and a set $D$ of edge attribute vectors that indicate the edge direction of nodes $u, v \in V$.
\end{definition}

The model that extracts the node features and edge attributes belong to the feature categories \circled{1}--\circled{5} is a netlist feature model.

\smallskip
Now, we can state the following theorem:
\begin{theorem}
    \label{thr:geaug_can_represent}
    The representation capability of a netlist feature extraction model is strictly greater than the representation capability of a baseline extraction model.
\end{theorem}
\begin{IEEEproof}
    %
    Let $\funcfg^{(1)}$ and $\funcfg^{(2)}$ be graph encoder models that are assumed to be injective.
    Since the extracted features from the netlist feature extraction model contain the features from the baseline feature extraction model, $\forall u, v \in V, u \ne v, f_{\mathsf{g}2; \Lambda_{\mathsf{baseline}}(\vartheta)} (u) \ne f_{\mathsf{g}2; \Lambda_{\mathsf{baseline}}(\vartheta)} (v) \implies f_{\mathsf{g}1; \Lambda_{\mathsf{netlist}}(\vartheta)} (u) \ne f_{\mathsf{g}1; \Lambda_{\mathsf{netlist}}(\vartheta)} (v)$.
    Thus, $P_{(\vartheta)}(\Lambda_{\mathsf{netlist}}, \funcfg^{(1)}) \succeq P_{(\vartheta)}(\Lambda_{\mathsf{baseline}}, \funcfg^{(2)})$.
    Furthermore, by Propositions \ref{prop:f3:gnn_loops}, \ref{prop:f5:gnn_eaug}, and \ref{prop:f6-3:gnn_functional_behavior}, there exists a set of nodes that the netlist feature extraction model can distinguish while the baseline feature extraction model cannot.
    Thus, $P_{(\vartheta)}(\Lambda_{\mathsf{netlist}}, \funcfg^{(1)}) \succ P_{(\vartheta)}(\Lambda_{\mathsf{baseline}}, \funcfg^{(2)})$.
    Therefore, Theorem~\ref{thr:geaug_can_represent} holds.
\end{IEEEproof}
By Theorem~\ref{thr:geaug_can_represent}, such a feature extraction model represents the feature categories \circled{1}--\circled{5}.

It should be noted that the assumption that graph encoder models are injective is a realistic modeling approach for theoretical analysis.
For simplicity, we consider a neural network model as a matrix product.
In general, the weights of the neural network model are initialized with Gaussian random values.
Since the weight matrices are full-rank in most cases, the matrix product becomes injective.
However, a precise analysis is remained to be a future work.

\subsection{Summary of HT Features}
This section presents the five feature categories for GL as a way to \circled{1}--\circled{5} and bridges the gap between HT features and GL.
As a result, this section clarifies what HT features a graph encoder model captures by theoretical analysis.

In the existing HT detection methods such as \cite{conf/iolts/KuriharaT21} and \cite{8702462}, we must discover effective features for HT detection.
To consider the topology around each node in a netlist, we should carefully analyze the relationship between a node to another node.
Therefore, the search space for feature engineering reaches $\funcO(|V| \times |V|)$.
Alternatively, the proposed method automatically extracts the structural features by graph extraction models.
It is enough to define initial feature vectors that represent the characteristics of nodes well.
To explore such features, we analyze all nodes and extract representative features.
Therefore, the search space for feature engineering becomes $\funcO(|V|)$.
We will demonstrate through experiments in Section~\ref{sec:6_evaluation} that a graph encoder model automatically extracts HT features.

%% file: tex/5_proposed.tex
\section{Proposed Method}
\label{sec:5_proposed}


\subsection{Overview}

\begin{figure*}[t]
    \center
    \includegraphics[width=0.90\linewidth]{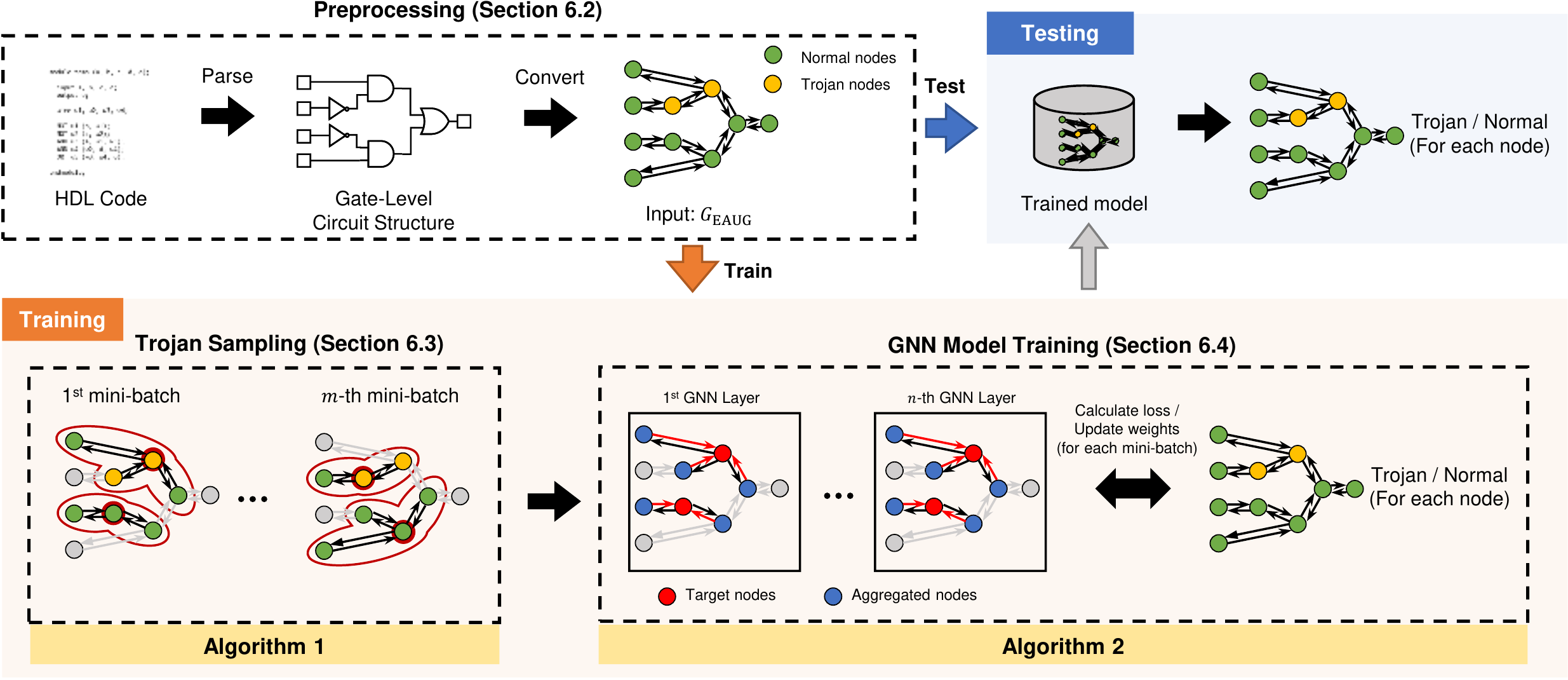}
    \caption{Overview of \tNAME.}
    \label{fig:overview}
\end{figure*}

Fig.~\ref{fig:overview} illustrates an overview of the proposed method, called \tNAME.
As shown in Fig.~\ref{fig:overview}, an HDL design is converted to a graph that represents the netlist, constructing an EAUG, $\Geaug$.
Training and/or testing of datasets are based on the EAUGs.

In the preprocessing phase (see Section~\ref{subsec:5_2_initial} in detail), we prepare netlists to be trained with the GNN model.
First, we convert the netlists to EAUG where each element of circuit is assigned to a node, and each wire is assigned to an edge (\circled{4}).
Then, we assign the initial feature value that covers the feature categories \circled{1}--\circled{3} and \circled{5} to each node in the graph.
At the same time, we assign the edge attribute that shows the direction of the edge.
Finally, we construct the dataset including multiple EAUGs.

In the training phase (see Section~\ref{subsec:5_3_trojan_sampling} and Section~\ref{subsec:5_4_gnn} in detail), we start from the Trojan sampling on the given EAUGs to balance the normal and Trojan nodes in each mini-batch.
For the training of HTs, we have to deal with a problem that HTs are quite tiny.
Because of the stealth of HTs, they are often constructed with a tiny scale.
Therefore, the numbers of genuine nodes and Trojan nodes are imbalanced in the training dataset.
To accurately perform \tNodeLevel classification for HT detection, we should adequately deal with the imbalanced dataset.
To address the problem, we propose the \textit{Trojan sampling} method to train the imbalanced HT dataset effectively.
After that, we train the mini-batches to classify each node in a graph as Trojan or normal.

In the testing phase, we just classify each node in an EAUG as either normal or Trojan using the trained model.


\subsection{Initial Feature Vector of a Node}
\label{subsec:5_2_initial}

\begin{table}[t]
    \centering
    \caption{Initial feature vector used in the proposed model\,\textsuperscript{a}.}
    \label{tbl:our-initial-node-features}
    \begin{tabular}{@{}p{\linewidth}@{}}
        \centering
        \begin{threeparttable}[t]
            \begin{tabular}{c|l|l} \hline
                No. & Feature & Category \\ \hline\hline
                1 & In-degree & \circled{1} \\
                2 & Out-degree & \circled{1} \\
                3--42 & Node types & \circled{2} \\
                43 & Min. distance to any primary input & \circled{3} \\
                44 & Min. distance to any primary output & \circled{3} \\
                45 & Static probability (0) of logic gates & \circled{5} \\
                46 & Static probability (1) of logic gates & \circled{5} \\ \hline
            \end{tabular}
            \begin{tablenotes}[para,flushleft,online,normal] 
                \item[a] We construct an EAUG, which is categorized as \circled{4}.
            \end{tablenotes}
        \end{threeparttable}
    \end{tabular}
\end{table}

Based on the discussion in Section~\ref{sec:4_ht_features}, we design the initial feature vector to be assigned to each node.

Table~\ref{tbl:our-initial-node-features} shows the initial feature vector assigned to each node $v$ in the proposed model.
Features 1 and 2, which are categorized as \circled{1}, correspond to the in-degree and out-degree of a node $v$, respectively.
To effectively train the features, the feature values are standardized.
Features 3--42, which are categorize as \circled{2}, show the types of each node $v$, e.g., a two-input AND gate or a flip-flop.
Features 43 and 44 show the minimum distance to any primary input and output, respectively.
Primary inputs and outputs are characteristic nodes for HT detection.
We use them as \textit{anchor-sets}, and thus the features are categorized as \circled{3}.
The features 45 and 46 show the probability that a logic gate outputs 0 and 1, respectively.
We give the standardized feature values to the initial vector.
They are helpful for calculating the \emph{functional behavior} of a circuit and can be categorized as \circled{5}.
In summary, the features cover the feature categories \circled{1}--\circled{3} and \circled{5} described in Section~\ref{sec:4_ht_features}.

\subsection{Trojan Sampling}
\label{subsec:5_3_trojan_sampling}
Since an HT is tiny compared to a normal circuit, the numbers of normal nodes and Trojan nodes are significantly imbalanced.
We should adequately address the issue to perform HT detection stably.
In the conventional ML models (not GL models), over-sampling (or under-sampling) approaches can be adopted to enhance the minority classes.
SMOTE~\cite{10.5555/1622407.1622416} is a well-known method for over-sampling, which synthetically generates minority class samples and enhances minority classes.
However, one cannot directly adopt the approach to graph-structured data.
A conventional over-sampling method cannot consider the adjacent matrix.
Although the graph-version SMOTE method has been proposed very recently~\cite{conf/wsdm/ZhaoZW21}, it is difficult to construct an accurate decoder model.
Therefore, in \tNAME, we first split normal nodes into several subgraphs.
Then, we construct mini-batches by combining a Trojan subgraph with different normal subgraphs.

\begin{figure}[t]
    \centering
    \scalebox{0.85}{
    \begin{minipage}{1.15\linewidth}
    \begin{algorithm}[H]
        \caption{Trojan Sampling Algorithm}
        \label{alg:trojan-sampling}
        \begin{algorithmic}[1]
            \REQUIRE Sets of Trojan and normal nodes $\setVt \cup \setVn = V$, the number of mini-batches $m$
            \ENSURE Set of mini-batches $\setB$
            \STATE $\setB \leftarrow \emptyset$
            \STATE Split $\setVn$ into $m$ subsets $\setVn^{(i)}$ randomly, where $i \in [m]$ and $|\setVn^{(i)}| \le \lceil |\setVn| / m \rceil$.
            \FOR {$i = 1$ to $m$}
                \STATE $B^{(i)} \leftarrow \{ \setVn^{(i)}, \setVt \}$
                \STATE $\setB \leftarrow \setB \cup \{ B^{(i)} \}$
            \ENDFOR
            \RETURN $\setB$
        \end{algorithmic}
    \end{algorithm}
    \end{minipage}
    }
\end{figure}

Algorithm~\ref{alg:trojan-sampling} shows the Trojan sampling algorithm.
Let $\setVt$ and $\setVn$ be a set of Trojan nodes and a set of normal nodes, respectively.
Given the number of mini-batches $m$, we first split the set of normal nodes into $m$ subsets so that $\setVn = \{ \setVn^{(1)}, \cdots, \setVn^{(m)} \}$.
Note that $|\setVn^{(i)}| \le \lceil |\setVn| / m \rceil$, where $|\cdot|$ shows the cardinality of a given set.
Then, we construct $m$ mini-batches $\setB = \{ B^{(1)}, \cdots, B^{(m)} \}$, where $B^{(i)} = \{ \setVn^{(i)}, \setVt \}, i \in [m]$.
We train mini-batches $\setB$ in each iteration.

\subsection{GNN Model}
\label{subsec:5_4_gnn}

As described in Section~\ref{sec:4_ht_features}, EAUG corresponds the feature category \circled{4}.
According to (\ref{eq:gnn-agg-edge}), we aggregate the edge attribute as well as the feature vectors of nodes adjacent to the node $v$.
We concatenate edge attributes and node feature vectors of $u$ for each $v$ in our implementation.

In order to classify each node $v$ as normal or Trojan, we assign a one-hot vector $\vecy_v \in \mathbb{R}^2$, where $(1, 0)$ shows normal and $(0, 1)$ shows Trojan.
Then, we can set the optimization problem using the loss function $\funcL$ as follows:
\begin{equation}
    \label{eq:gnn-loss}
    \min \sum_{\forall v \in V} \funcL \left( \vecy_v, f( \vecx_{v}) \right),
\end{equation}
where $f(\vecx_{v}) = \funcfc ( \funcfg (\vecx_{v}) ).$
We repeat the training for each graph $G$ in a training dataset.

\begin{figure}[t]
    \centering
    \scalebox{0.85}{
    \begin{minipage}{1.15\linewidth}
    \begin{algorithm}[H]
        \caption{\tNAME: Training Algorithm}
        \label{alg:eaug-training}
        \begin{algorithmic}[1]
            \renewcommand{\algorithmiccomment}[1]{\bgroup\hfill//~#1\egroup}
            \REQUIRE Set of EAUGs $\mathcal{G}_\mathsf{EAUG}$, set of labels $Y$, the number of mini-batches $m$
            \ENSURE Trained model $f$
            \STATE Initialize the model $f$.
            \REPEAT[Repeat one-epoch process]
                \FORALL {$G_\mathsf{EAUG} = (V, E, X, D) \in \mathcal{G}_\mathsf{EAUG}$}
                    \STATE $\mathcal{B} \leftarrow$ \textbf{TrojanSamplingAlgorithm}$(V, m)$
                    \FORALL {$B \in \mathcal{B}$}
                        \STATE Calculate the loss of the model $f$ with $\vecx_{v}$, $\vecd_{\funcN(v)\rightarrow v}$, and $\vecy_{v}$ for all $v \in B$.
                        \STATE Update the weight of the model $f$.
                    \ENDFOR
                \ENDFOR
            \UNTIL{Training converges.}
            \RETURN $f$
        \end{algorithmic}
    \end{algorithm}
    \end{minipage}
    }
\end{figure}

Algorithm~\ref{alg:eaug-training} describes the training algorithm.
Let $\mathcal{G}_\mathsf{EAUG}$ be a set of EAUGs that corresponds to a set of netlists to be trained.
$Y$ is a set of labels $\vecy_{v}$ assigned to each node $v$.
We draw mini-batches $\setB$ from each EAUG, $G_\mathsf{EAUG}$ and calculate the loss of the model $f$ like (\ref{eq:gnn-loss}) using the nodes in a mini-batch.
Then, we update the weight of the model $f$ so as to minimize the loss between the model's outputs and the labels $\vecy_{v}, v \in B$.
We repeat the one-epoch process (ll. 3--9 in Algorithm~\ref{alg:eaug-training}) several times until the training converges.
In other words, we end the training process when the number of processed epochs reaches the limit, or the loss of the model $f$ no longer decreases.

%% file: tex/6_evaluation.tex
\section{Evaluation}
\label{sec:6_evaluation}

The evaluation in this section aims to answer the following research questions through the experiments:
\begin{itemize}
    \item \textbf{RQ1}: Does the domain knowledge on HTs enhance the detection performance? (Section~\ref{subsec:6_2_comparison}-a)
    \item \textbf{RQ2}: Does the proposed method enhance the detection performance for node-wise HT detection in netlists? (Section~\ref{subsec:6_2_comparison}-b)
    \item \textbf{RQ3}: Does the GL automatically extract effective features? (Section~\ref{subsec:6_3_unknown_features})
\end{itemize}

Each research question corresponds to the motivations described in Section~\ref{sec:3_motivation}: \textbf{RQ1} corresponds to \tMotivationC, \textbf{RQ2} corresponds to \tMotivationA, and \textbf{RQ3} corresponds to \tMotivationB.

\subsection{Setup}
\label{subsec:6_1_setup}
\noindent
\textbf{Datasets}.
For the dataset, we use 24 netlists from Trust-HUB~\cite{shakya2017benchmarking,Salmani2013} (the detail is presented in Appendix~\ref{apdx:benchmarks-metrics}).
In the Trust-HUB benchmarks, HTs with various structures are embedded into several types of netlists, and the insertion points are indicated by comments in the source code.
%
The total number of nodes in the dataset is 621140, of which the number of Trojan nodes is 1262. This means that Trojan nodes are only 0.2\% of the total data. The Trojan Sampling Algorithm described in Section~\ref{subsec:5_3_trojan_sampling} is applied to better learn imbalanced training data.

Furthermore, randomly generated samples are employed in Section~\ref{subsec:6_3_unknown_features}.
The details are presented in the section.

\noindent
\textbf{Models}.
We train and identify the dataset with the graph structure described in the {Datasets} section.
We use GAT~\cite{conf/iclr/VelickovicCCRLB18}, MPNN~\cite{10.5555/3305381.3305512}, and GIN~\cite{conf/iclr/XuHLJ19} as the GNN models for the evaluation.
The graph encoder models for EAUGs are described in Appendix~\ref{apdx:gnn_for_eaug}.
We set the following parameters for all the models.
The maximum number of epochs is 1000, and early stopping~\cite{yao2007early} is applied, which finishes the training process when the loss does not decrease for 50 epochs. The learning rate is 0.1, the optimization algorithm is Adam, the activation function is an exponential linear unit~(ELU) function, and binary cross-entropy is used for the loss function.

Since the best parameters for the number of mini-batches~(\#batches), the number of GNN layers~(\#layers), and the number of dimensions of the feature vectors after the graph convolution~(\#units) depend on the GNN model, we search for them using grid search. 
We explain the details in Appendix~\ref{apdx:gnn_parameters}.
Table~\ref{tbl:best-param} shows the best parameters in each model for the proposed method and the baseline method described in Section~\ref{subsec:6_2_comparison}.
\begin{table}[t]
    \centering
    \caption{Best parameters for each model.}
    \label{tbl:best-param}
    \begin{tabular}{c|c||ccc} \hline
    Method                    & Model & \#batches & \#layers & \#units \\ \hline \hline
    \multirow{3}{*}{Proposed} & GAT   & 20      & 3       & 16     \\
                              & MPNN  & 15      & 2       & 32     \\
                              & GIN   & 20      & 2       & 16     \\ \hline
    \multirow{3}{*}{Baseline} & GAT   & 5       & 3       & 32     \\ 
                              & MPNN  & 15      & 3       & 32     \\
                              & GIN   & 20      & 2       & 32     \\ \hline
    \end{tabular}
\end{table}


\noindent
\textbf{Evaluation metrics}.
To evaluate the performance, we perform a leave-one-out cross-validation. For each of the 24 netlists described in the Datasets section, we use one as a test sample and the remaining 23 as training samples. We perform this validation on all the netlists and evaluate the average of the 24 classification results.
The evaluation metrics are recall, precision, F1-score, and accuracy, in which a Trojan net is regarded as a positive sample.

\subsection{Performance Evaluation for Trust-HUB Benchmarks Comparing with Baseline and  State-of-the-Art Methods}
\label{subsec:6_2_comparison}

\subsubsection{Comparison with the baseline method (\textbf{RQ1})}
In this experiment, the performance with and without domain knowledge in the features is evaluated.
As mentioned in Section~\ref{subsec:3_3_mot3}, feature selection based on domain knowledge is effective in GNNs.
To confirm this assumption, we compare the features of the proposed method as mentioned in Table~\ref{tbl:our-initial-node-features} with the features consisting only of node types.
In general, the node type is the most primitive feature when representing the information in a graph format \cite{muralidhar2021contrastive}.
Therefore, we pick up features 3--42, which are categorized as \circled{2} as shown in Table~\ref{tbl:our-initial-node-features}.
We define them as baseline node features that do not contain domain knowledge and GL based on them is considered to be the {\it baseline method}. 

To compare the results fairly, we search for the best parameters of the baseline method by grid search as in the proposed method. The details are shown in Appendix~\ref{apdx:gnn_parameters}.


\noindent
\textbf{Detection results}.
Table~\ref{tbl:res-cmp-baseline} shows the detection results of this experiment (the detailed results are shown in Appendix~\ref{apdx:best_parameter}).
The boldfaced font indicates the highest rate in each method.
As shown in Table~\ref{tbl:res-cmp-baseline}, the proposed method performed as well or better than the baseline method in all evaluation metrics.
In particular, the GAT model outperforms the other models in all metrics. Therefore, GAT is the best GNN model for HT detection in Trust-HUB.

This means that we can say, ``YES'' to \textbf{RQ1} corresponding to \tMotivationC.
By eliminating the feature categories \circled{1}, \circled{3}, \circled{4}, and \circled{5}, which are based on the domain knowledge of the HTs, the recall, precision and F1-score are decreased.
The accuracy is the same as the baseline method because the Trojan node is only 0.2\% of the whole training data, and the accuracy score is greatly affected by true negatives.
For example, if the classifier determines all nodes as normal nets (i.e., negative), the accuracy would be 99.8\%.
Therefore, there is no difference in accuracy among the methods, but it is clear that the proposed method can identify the Trojan nodes better than the baseline method because the recall, precision, and F1-score are higher.
This confirms that the features based on knowledge of HT features are essential for node-wise HT detection.
\begin{table}[t]
    \centering
    \caption{Detection results of Trust-HUB for the proposed method and the baseline method.}
    \label{tbl:res-cmp-baseline}
    \resizebox{\linewidth}{!}{
    \begin{tabular}{c|c||cccc} \hline
    Method                    & Model & Recall & Precision & F1-score & Accuracy \\ \hline \hline
    \multirow{3}{*}{Proposed} & GAT   & \textbf{0.890}  & \textbf{0.978}     & \textbf{0.921}    & \textbf{0.998}    \\
                              & MPNN  & 0.865  & 0.933     & 0.887    & \textbf{0.998}    \\
                              & GIN   & 0.585  & 0.616     & 0.498    & 0.988    \\ \hline
    \multirow{3}{*}{Baseline} & GAT   & \textbf{0.880}  & \textbf{0.976}     & \textbf{0.915}    & \textbf{0.998}    \\
                              & MPNN  & 0.879  & 0.900     & 0.871    & \textbf{0.998}    \\
                              & GIN   & 0.237  & 0.392     & 0.240    & 0.986    \\ \hline
    \end{tabular}
    }
\end{table}

\subsubsection{Comparison with state-of-the-art methods (\textbf{RQ2})}
In this experiment, we compare the proposed method with the two state-of-the-art methods~\cite{conf/iolts/KuriharaT21,8702462} mentioned in Section~\ref{subesc:4_1_HT_features}.

The method of \cite{conf/iolts/KuriharaT21} extracts 36 features that represent the structure of HT well, and identifies HT wires by the random forest~(RF).
Since the number of false positives is very small, this method is less likely to misidentify normal circuits.
The method of \cite{8702462} is based on testability measures, which are effective features for HT detection, and employs ADASYN to learn imbalanced training data better.
They validated it with four supervised algorithms, with the highest metrics when using the bagged trees~(BT).
Both methods \cite{conf/iolts/KuriharaT21,8702462} have been reported to perform well on the Trust-HUB dataset based on the effective feature engineering.

\noindent
\textbf{Detection results}.
Table~\ref{tbl:res-cmp-exist} shows the comparison results with existing methods.
We adopt the GAT model, which shows the highest classification performance as shown in Table~\ref{tbl:res-cmp-baseline}, as the proposed method for the comparison.
As shown in Table~\ref{tbl:res-cmp-exist}, the proposed method outperforms the two methods~\cite{conf/iolts/KuriharaT21,8702462} in all evaluation metrics.

This means that we can say, ``YES'' to \textbf{RQ2} corresponding to \tMotivationA.
As far as we know, there is no other gate-level and node-wise HT detection method that achieves 0.890 recall and 0.978 precision.
Since we can accurately identify the insertion point of the HT, we can obtain the evidence for further analysis.
Moreover, because of the high accuracy, we can carefully analyze or refine the suspicious circuit using other verification techniques based on the classification results of the proposed method.
Therefore, the proposed method solves the problem in practical use.
\begin{table}[t]
    \centering
    \caption{Comparison with existing methods.}
    \label{tbl:res-cmp-exist}
    \resizebox{\linewidth}{!}{
    \begin{tabular}{c|c||cccc} \hline
    Method   & Model & Recall & Precision & F1-score & Accuracy \\ \hline \hline
    Proposed & GAT   & \textbf{0.890}  & \textbf{0.978}     & \textbf{0.921}    & \textbf{0.998}    \\ \hline
    \cite{conf/iolts/KuriharaT21} & RF    & 0.636  & 0.957     & 0.667    & 0.994    \\
    \cite{8702462}    & BT    & 0.825  & 0.866     & 0.827    & 0.983  \\ \hline
    \end{tabular}
    }
\end{table}

\subsection{Performance Evaluation for HTs with Unknown Features (\textbf{RQ3})}
\label{subsec:6_3_unknown_features}

In this experiment, we evaluate the proposed method for HTs with unknown features using randomly generated samples.
As mentioned in Section~\ref{subsec:3_2_mot2}, GNN models are expected to automatically extract the features that represent HTs well.
Although conventional ML-based methods require feature engineering to effectively perform HT detection, GNNs will overcome it.
To confirm this assumption, we randomly generate HT-infested samples and attempt to detect HTs from the generated samples without feature engineering of their HTs.
For comparison, we adopt the \textit{baseline method} and the method of \cite{conf/iolts/KuriharaT21} in this experiment.

\noindent
\textbf{Random HT-infested circuit generation}.
Inspired by methodology in \cite{conf/date/CruzHMB18}, HT-infested circuits for the evaluation were randomly generated.
How to randomly generate HT-infested circuits is described in Appendix~\ref{apdx:random_ht_generation}.

We randomly generate 20 training samples, including 79155 normal nodes and 2227 Trojan node, and  100 test samples, including 396103 normal nodes and 9961 Trojan nodes.
In the generated HTs, most Trojan nodes are less than 5\% of the total number of nodes.
The generated HTs are relatively small compared to the normal circuits.
For evaluation, we train the 20 training samples and construct a trained model.
Then, we classify 100 test samples and calculate the average scores of all of the samples in terms of recall, precision, F1-score, and accuracy.


In this experiment, we use the same model as in the previous section.
Specifically, we use the same parameters as in the previous section for the evaluation of the proposed method and the baseline method.
We implement the method in \cite{conf/iolts/KuriharaT21} for comparison with a state-of-the-art HT detection method, which considers the feature categories \circled{1}--\circled{4} and is designed based on the features that appear in the Trust-HUB benchmark netlists.


\noindent
\textbf{Detection results}.
Table~\ref{tbl:res-random-gen} shows the detection results of this experiment.
The boldfaced font indicates the highest rate in each method.
As shown in Table~\ref{tbl:res-random-gen}, the GNN-based methods (the proposed method and the baseline method) outperform the method in \cite{conf/iolts/KuriharaT21} in any evaluation metrics.
Although the precision of \cite{conf/iolts/KuriharaT21} is higher than the proposed method in Table~\ref{tbl:res-cmp-exist}, it was not as high in this experiment.
Instead, all the metrics are quite low compared to the results of GNN models.
This is because the features in \cite{conf/iolts/KuriharaT21} are designed for only the Trust-HUB benchmark netlists, and thus this method fails to capture the features of randomly generated HTs (i.e., the HTs with unknown features).
On the other hand, the GNN-based methods successfully capture the node features of HTs and achieve higher detection performance.
From the results, we can state that GNN-based HT detection successfully extracts HT features without tedious feature engineering, even when the datasets are not involved in a specific benchmark suite.
This means that we can say, ``YES'' to \textbf{RQ3} corresponding to \tMotivationB.

\begin{table}[t]
    \centering
    \caption{Detection results of unknown HTs.}
    \label{tbl:res-random-gen}
    \begin{tabular}{c|c||cccc} \hline
        Method & Model & Recall & Precision & F1-score & Accuracy \\ \hline \hline
        \multirow{3}{*}{Proposed} & GAT & 0.773 & 0.843 & 0.788 & 0.995 \\ 
          & MPNN & \textbf{0.868} & \textbf{0.993} & \textbf{0.905} & \textbf{0.997} \\ 
          & GIN & 0.783 & 0.847 & 0.800 & 0.995 \\ \hline
        \multirow{3}{*}{Baseline} & GAT & 0.756 & 0.889 & 0.781 & 0.995 \\ 
          & MPNN & 0.854 & 0.936 & 0.881 & 0.996 \\ 
          & GIN & \textbf{0.863} & \textbf{0.965} & \textbf{0.897} & \textbf{0.997} \\ \hline
        \cite{conf/iolts/KuriharaT21} & RF & 0.681 & 0.749 & 0.706 & 0.996 \\ \hline
    \end{tabular}
\end{table}


%% file: tex/7_conclusion.tex
\section{Conclusion}
\label{sec:7_conclusion}



In this paper, a novel HT detection method in netlists using GL called \tNAME was proposed.
\tNAME applies node-wise detection in netlists, GL, and domain knowledge of HTs for practical use.
Thus, this paper theoretically supports the relationship between GL and HT detection and clarifies what HT features GL captures.
Based on the theoretical analysis described in Section~\ref{sec:4_ht_features}, it is established that \tNAME effectively captures the HT features.
The experimental results demonstrate that \tNAME successfully outperforms the existing HT detection methods and extracts HT features without tedious feature engineering.

%% file: tex/8_appendix.tex
\appendices

\section{Proof of Proposition~\ref{prop:f3:gnn_loops}}
\label{apdx:prf:prop:f3:gnn_loops}

\begin{IEEEproof}
    Let $\xi(v, \mathcal{A})$ be a function that calculates the minimum distance from a node $v$ to any node in an anchor-set $\mathcal{A}$.
    The node feature vector of a node $v$ at the $l$-th layer is represented as $\{\vechl_{v}, \xi(v, \mathcal{A}) \}$.
    Then, according to (\ref{eq:gnn-agg}), the feature vectors of $v$ and $u$ at the first layer are expressed as follows:
    \begin{align}
        \vecm^{(1)}_{v} & = \mathsf{AGGREGATE}^{(1)} \left( \left\{ \{ \vecx_{v'}, \xi(v', \mathcal{A}) \}: \forall v' \in \mathcal{N}(v) \right\} \right), \nonumber \\
        \vecm^{(1)}_{u} & = \mathsf{AGGREGATE}^{(1)} \left( \left\{ \{ \vecx_{u'}, \xi(u', \mathcal{A}) \}: \forall u' \in \mathcal{N}(u) \right\} \right). \nonumber
    \end{align}
    If $\mathsf{AGGREGATE}^{(l)}, \forall l \in [L]$ is injective, $\xi(v, \mathcal{A}) \ne \xi(u, \mathcal{A})\implies \vecm^{(l)}_{v} \ne \vecm^{(l)}_{u}$.
    By recursively processing (\ref{eq:gnn-agg}) and (\ref{eq:gnn-comb}), $\vecz_{v} \ne \vecz_{u}$.
    Thus, the nodes $u$ and $v$ are distinguishable, and Proposition \ref{prop:f3:gnn_loops} holds.
\end{IEEEproof}

\section{Proof of Proposition~\ref{prop:f5:gnn_eaug}}
\label{apdx:prf:prop:f5:gnn_eaug}

\begin{IEEEproof}
    We assign the edge attributes as $\vecd_{u \rightarrow v} \ne \vecd_{v \rightarrow u}: \forall v \in V, \forall u \in \mathcal{N}(v), u \ne v$.
    Then, \circleedged{c} can be represented obviously.
    Since an EAUG inherits the properties of an undirected graph, \circleedged{d} can also be represented.
    Here we modify (\ref{eq:gnn-agg}) to involve edge attributions.
    \begin{equation}
        \vecm^{(l+1)}_{v} = \mathsf{AGGREGATE}^{(l)} \left( \left\{ \vechl_{u}, \vecd_{u \rightarrow v}: \forall u \in \mathcal{N}(v) \right\} \right) \label{eq:gnn-agg-edge}
    \end{equation}
    If the aggregate function $\mathsf{AGGREGATE}^{(l)}(\cdot)$ distinguishes the edge attributes, the aggregated message $\vecm^{(l+1)}_{v}$ implicitly involves both input and output side structures.
    Since $\mathsf{AGGREGATE}^{(l)}(\cdot)$ is injective in the well-trained graph encoder model, the input and output sides can be distinguished.
    As a result, the downstream task can distinguish \circleedged{a} and \circleedged{b}.
    Therefore, the proposition holds.
\end{IEEEproof}

\section{Proof of Proposition~\ref{prop:f6-3:gnn_functional_behavior}}
\label{apdx:prf:prop:f6-3:gnn_functional_behavior}

\begin{IEEEproof}
    Let $z_{\theta}$ be a learnable function parameterized by $\theta$.
    Here, we express the function to update the hidden feature vector of a node $v$ at the $l$-th layer of a graph encoder model $\funcfg$ as follows:
    \begin{equation}
        \label{eq:gate-func-comb}
        \vech^{(l+1)}_{v} = z_{\theta^{(l)}} \left( \vechl_{v}, \tilde{\vech}^{(l)}_{v} \right),
    \end{equation}
    where $\theta^{(l)}$ is a parameter of the function $z$ to be an injective function, and $\tilde{\vech}^{(l)}_{v}$ is an aggregated feature vector of adjacent nodes that is calculated as follows:
    \begin{equation}
        \label{eq:gate-func-agg}
        \tilde{\vech}^{(l)}_{v} = \mathsf{AGGREGATE}^{(l)} \left( \left\{ \vechl_{u}: \forall u \in \mathcal{N}(v) \right\} \right)
    \end{equation}
    Now we assign a node type $\funcType(v)$ and a feature value of a functional behavior $\funcBehavior(v)$ to a node $v$.
    Then, (\ref{eq:gate-func-comb}) can express (\ref{eq:functional_behavior}).

    If let $z_{\theta^{(l)}}(\cdot, \cdot)$ be $\mathsf{COMBINE}^{(l)}$ and $\tilde{\vech}^{(l)}_{v}$ be $\vecm^{(l)}_{v}$, the equations (\ref{eq:gate-func-comb}) and (\ref{eq:gate-func-agg}) correspond to the equations (\ref{eq:gnn-comb}) and (\ref{eq:gnn-agg}), respectively.
    Thus, there exists a graph encoder model with the function $z_{\theta^{(j)}}$ that identifies the \emph{functional behavior} within $k$ hops from each node $v$ for a given $k$.

    In terms of the direction of calculating \emph{functional behavior}, it is shown that an EAUG identifies edge directions by Proposition~\ref{prop:f5:gnn_eaug}.
    This means that a graph encoder model exists that identifies \emph{functional behaviors} with distinguishing the input and output directions.

    Hence, the proposition holds.
\end{IEEEproof}

\section{Benchmarks}
\label{apdx:benchmarks-metrics}

Table~\ref{tbl:benchmarks} shows the benchmark netlists used in the experiments in Section~\ref{subsec:6_2_comparison}.

\begin{table}[t]
    \centering
    \caption{Benchmarks.}
    \label{tbl:benchmarks}
    \begin{tabular}{c|cc} \hline
    Netlist              & \begin{tabular}{c}Normal \\ nodes\end{tabular} & \begin{tabular}{c}Trojan \\ nodes\end{tabular} \\ \hline\hline
    RS232-T1000          & 289    & 13     \\
    RS232-T1100          & 293    & 11     \\
    RS232-T1200          & 296    & 10     \\
    RS232-T1300          & 290    & 9      \\
    RS232-T1400          & 290    & 12     \\
    RS232-T1500          & 291    & 13     \\
    RS232-T1600          & 290    & 13     \\
    s15850-T100          & 2397   & 27     \\
    s35932-T100          & 5967   & 15     \\
    s35932-T200          & 5962   & 15     \\
    s35932-T300          & 5975   & 26     \\
    s38417-T100          & 5656   & 12     \\
    s38417-T200          & 5656   & 15     \\
    s38417-T300          & 5688   & 15     \\
    s38584-T100          & 7064   & 9      \\
    s38584-T200          & 7064   & 83     \\
    s38584-T300          & 7064   & 731    \\
    EthernetMAC10GE-T700 & 102453 & 13     \\
    EthernetMAC10GE-T710 & 102453 & 13     \\
    EthernetMAC10GE-T720 & 102453 & 13     \\
    EthernetMAC10GE-T730 & 102453 & 13     \\
    B19-T100             & 63170  & 83     \\
    B19-T200             & 63170  & 83     \\
    wb\_conmax-T100      & 23194  & 15     \\ \hline
    Total                & 619878 & 1262   \\ \hline
    \end{tabular}
\end{table}


\section{GNN Models for EAUGs}
\label{apdx:gnn_for_eaug}
We show the models used in this paper as well as the layer definition in the PyTorch Geometric library~\cite{Fey/Lenssen/2019}.

\smallskip
\noindent
\textsl{GAT}:
The original \textsl{GAT} convolution layer is defined as \texttt{GATConv} in PyTorch Geometric.
Based on the \texttt{GATConv} layer, we extend the methods to adopt to an EAUG.
The \textsl{GAT} model used in this paper is expressed as follows:
\begin{equation}
    \begin{split}
        \vecm^{(l)}_{v} & = \sum\nolimits_{u \in \mathcal{N}(v)} \alpha^{(l)}_{u, v} \matWl \cdot [ \vechl_v \, \Vert \, \vecd_{u \rightarrow v} ] \\
        \vech^{(l+1)}_{v} & = \alpha^{(l)}_{v, v} \matWl \vechl_{v} + \vecm^{(l)}_{v}, \label{eq:gat-eaug}
    \end{split}
\end{equation}
where $[\cdot \, \Vert \, \cdot]$ concatenates given vectors and $\matWl$ is the parameter of the shared linear transformation.
$\alpha^{(l)}_{u, v}$ is an attention to an edge from a node $u$ to another node $v$ at the $l$-th layer:%
\begin{equation}
    \alpha^{(l)}_{u, v} = \frac{ \exp{ \left( \funcACT \left( \funcH \left( \matWl \vechl_v, \matWl \vechl_u \right) \right) \right)} }{ \sum_{w \in \mathcal{N}(v)} \exp{ \left( \funcACT \left( \funcH \left(\matWl \vechl_v, \matWl \vechl_w \right) \right) \right) } }, \label{eq:gat-eaug-att}
\end{equation}
where $\funcH(\cdot, \cdot)$ is a learnable function, and $\funcACT$ is an activation function.
We use $\mathrm{LeakyReLU}$ as the activation function $\funcACT$.

\smallskip
\noindent
\textsl{MPNN}:
This convolution layer is defined as \texttt{NNConv} in  PyTorch Geometric.
The \textsl{MPNN} model used in this paper is expressed as follows:
\begin{equation}
    \begin{split}
        \vecm^{(l)}_{v} & = \sum\nolimits_{u \in \mathcal{N}(v)} \vechl_u \cdot \funcMLP_{\theta} \left( \vecd_{u \rightarrow v} \right) \\
        \vech^{(l+1)}_{v} & = \matWl \vechl_{v} + \vecm^{(l)}_{v}, \label{eq:mpnn-eaug}
    \end{split}
\end{equation}
where $\funcMLP_{\theta}$ is an MLP parameterized by $\theta$ and $\matWl$ is a weight matrix.

\smallskip
\noindent
\textsl{GIN}:
To adopt to the edge attributes in an EAUG, we use the model leveraged in \cite{conf/iclr/HuLGZLPL20}.
This convolution layer is defined as \texttt{GINEConv} in PyTorch Geometric.
The \textsl{GIN} model used in this paper is expressed as follows:
\begin{equation}
    \begin{split}
        \vecm^{(l)}_{v} & = \sum\nolimits_{u \in \mathcal{N}(v)} \mathrm{ReLU} \left( \vechl_u + \matWl \vecd_{u \rightarrow v} \right) \\
        \vech^{(l+1)}_{v} & = \funcMLP_{\theta} \left( \left( 1 + \epsilon^{(l)} \right) \cdot \vechl_v + \vecm^{(l)}_{v} \right), \label{eq:gin-eaug}
    \end{split}
\end{equation}
where $\matWl$ is a weight matrix and $\funcMLP_{\theta}$ is an MLP parameterized by $\theta$.

\section{GNN Parameters}
\label{apdx:gnn_parameters}
For the experiment in Section~\ref{subsec:6_2_comparison}, we conduct a grid search to find the best parameters for each model, GAT, MPNN, and GIN. 
We vary the number of mini-batches, the number of GNN layers, and the dimensions of the feature vectors updated by the aggregation.
We summarize the parameters in Table~\ref{tbl:parameters}.
We compare the recall, precision, and F1-score of the classification results, and select the model that shows the highest value in two or more evaluation metrics as the best parameter.
When several models show similar performance, we focus on F1-score to make a decision because we aim to maximize both recall and precision as much as possible.

\begin{figure*}[t]
\centering
    \begin{minipage}[t]{0.32\linewidth}
        \centering
        \includegraphics[keepaspectratio, scale=0.22]{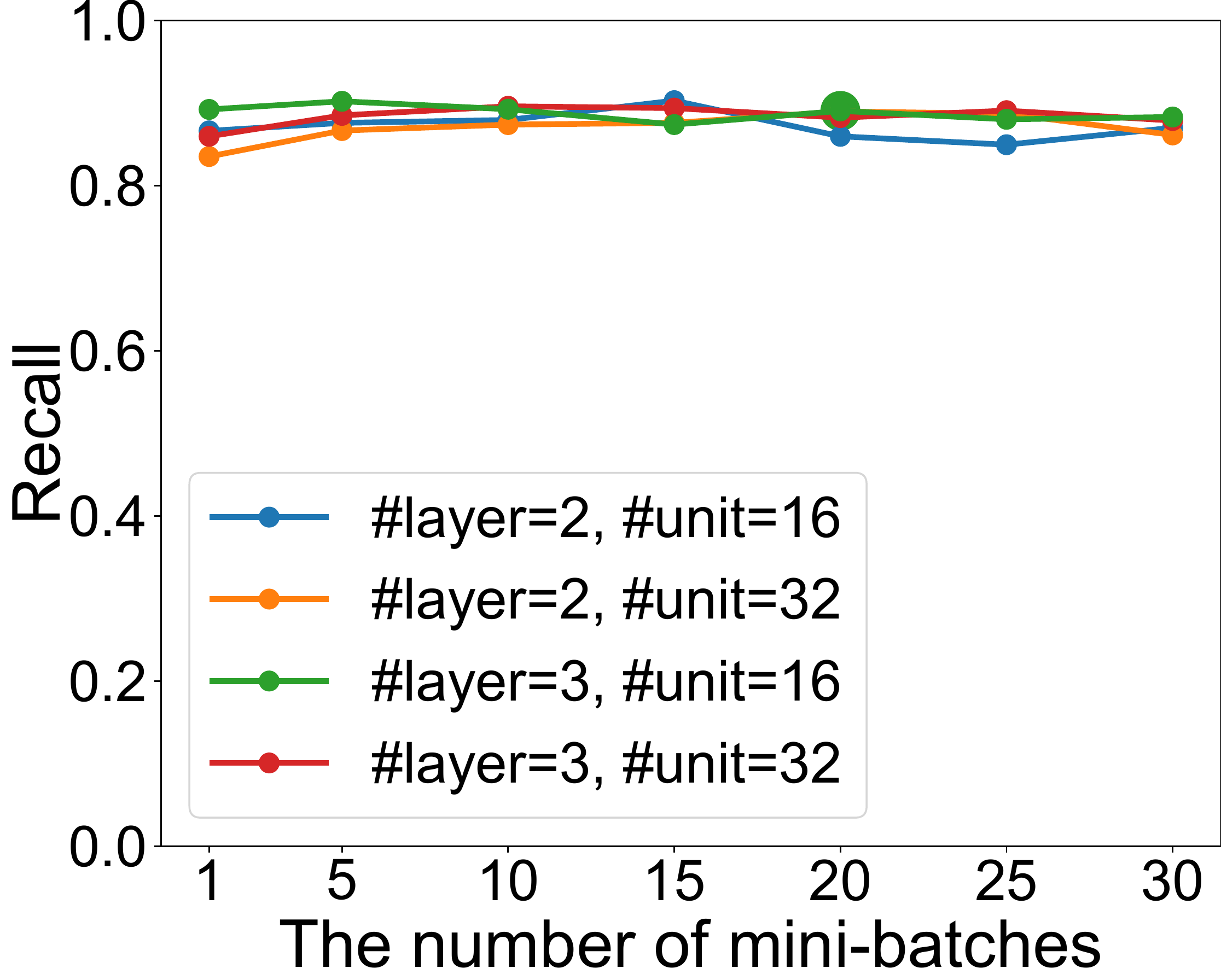}
        \subcaption{Recall of GAT.}
    \end{minipage}
    \begin{minipage}[t]{0.32\linewidth}
        \centering
        \includegraphics[keepaspectratio, scale=0.22]{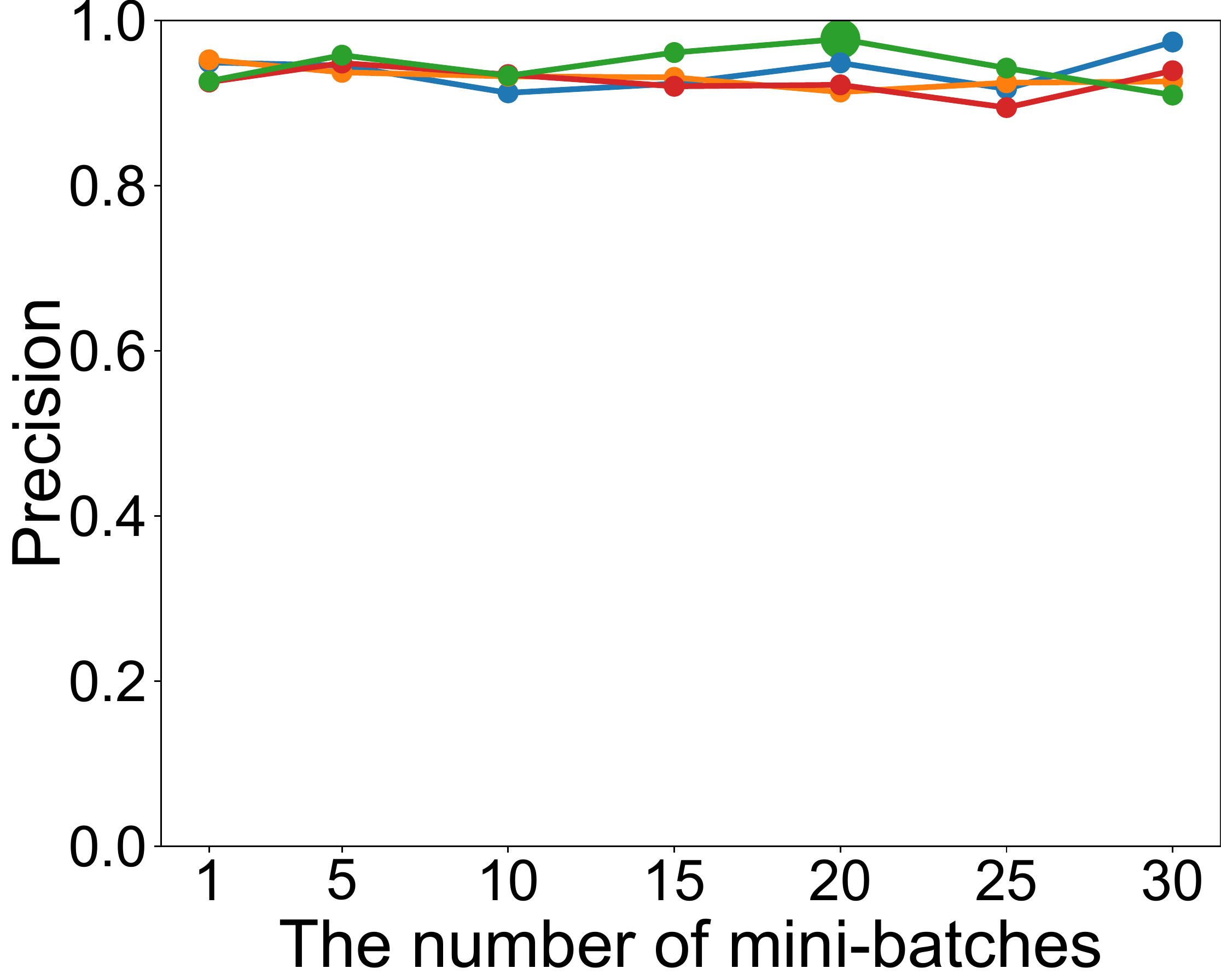}
        \subcaption{Precision of GAT.}
    \end{minipage}
    \begin{minipage}[t]{0.32\linewidth}
        \centering
        \includegraphics[keepaspectratio, scale=0.22]{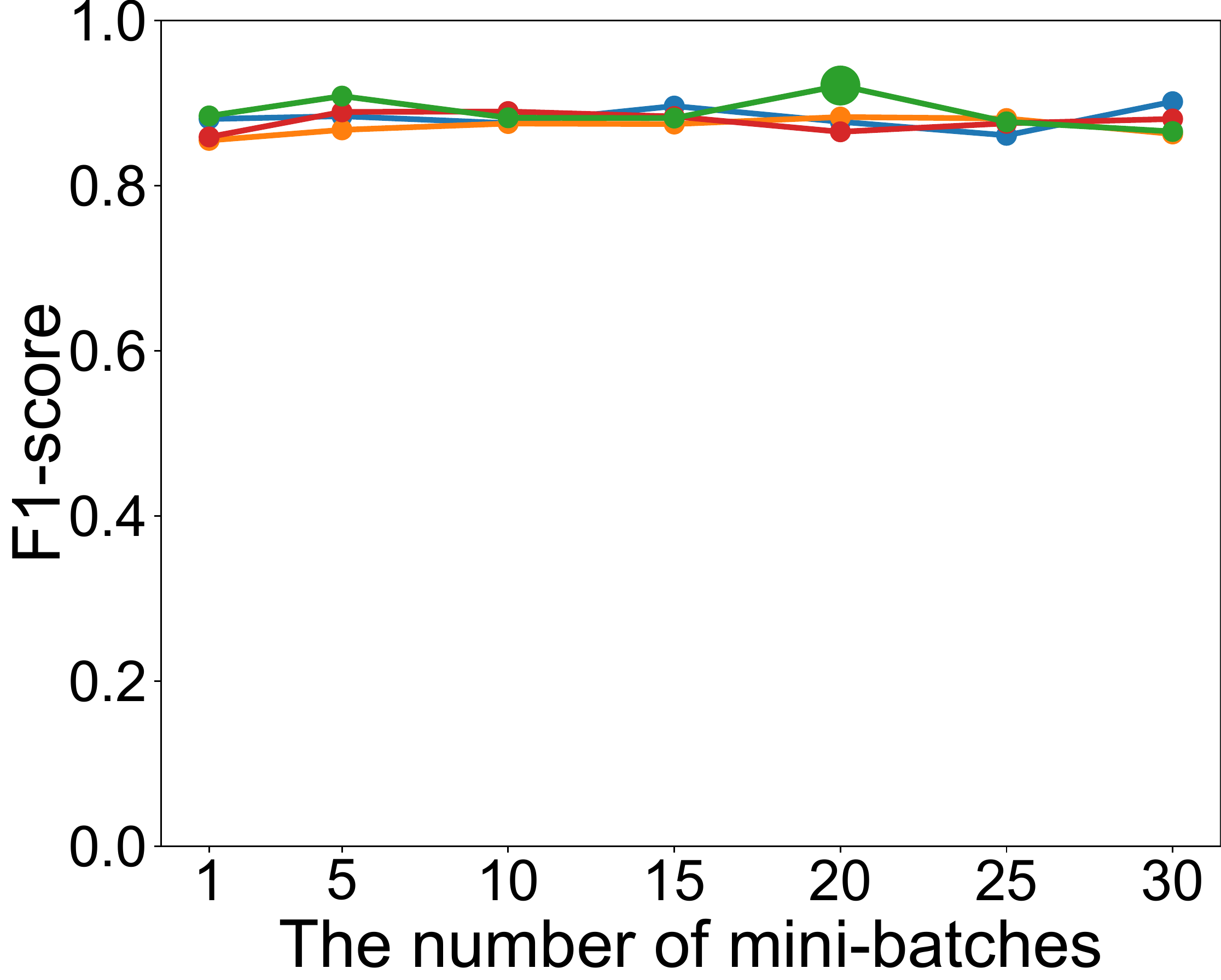}
        \subcaption{F1-score of GAT.}
    \end{minipage}\\
    \begin{minipage}[t]{0.32\linewidth}
        \centering
        \includegraphics[keepaspectratio, scale=0.22]{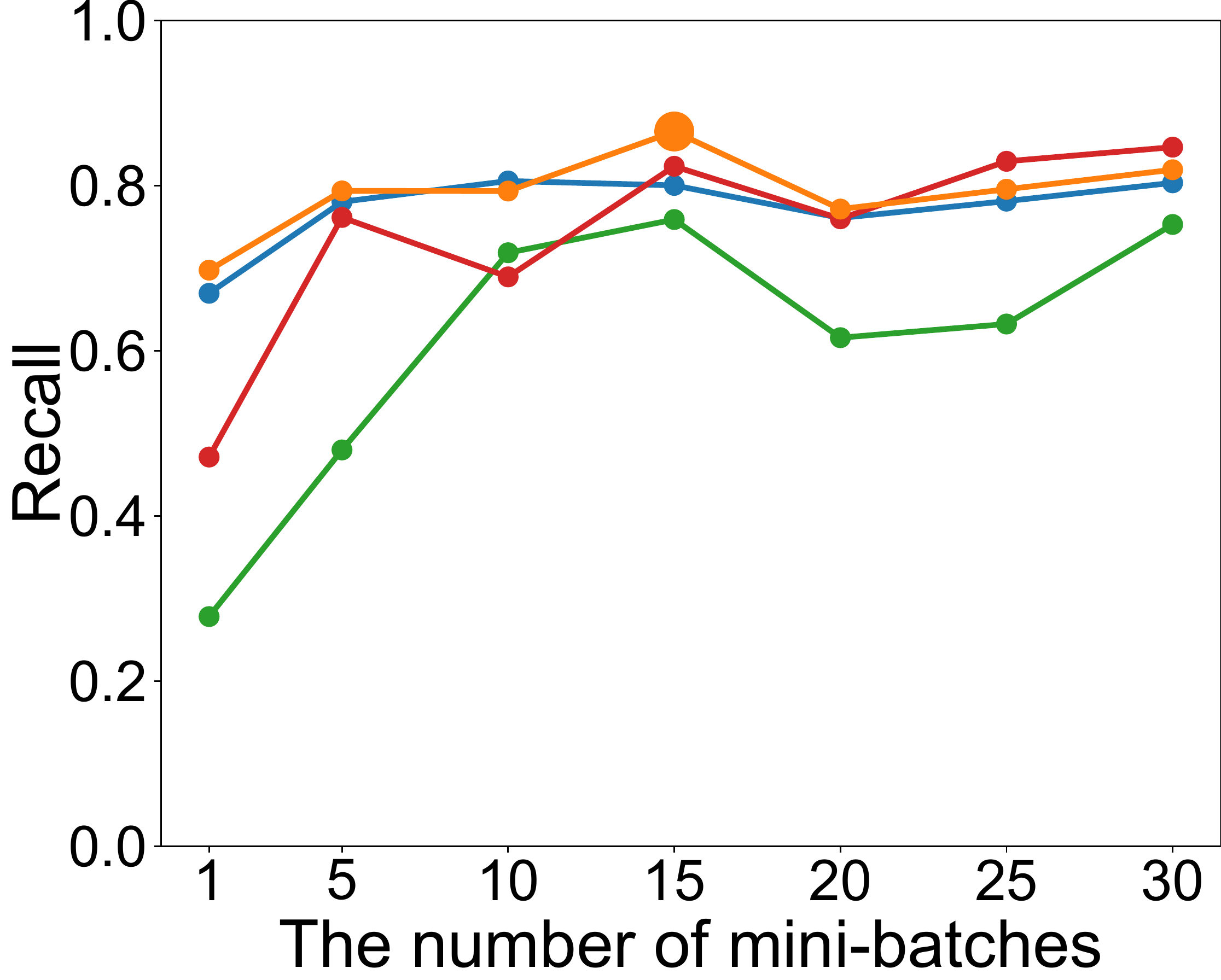}
        \subcaption{Recall of MPNN.}
    \end{minipage}
    \begin{minipage}[t]{0.32\linewidth}
        \centering
        \includegraphics[keepaspectratio, scale=0.22]{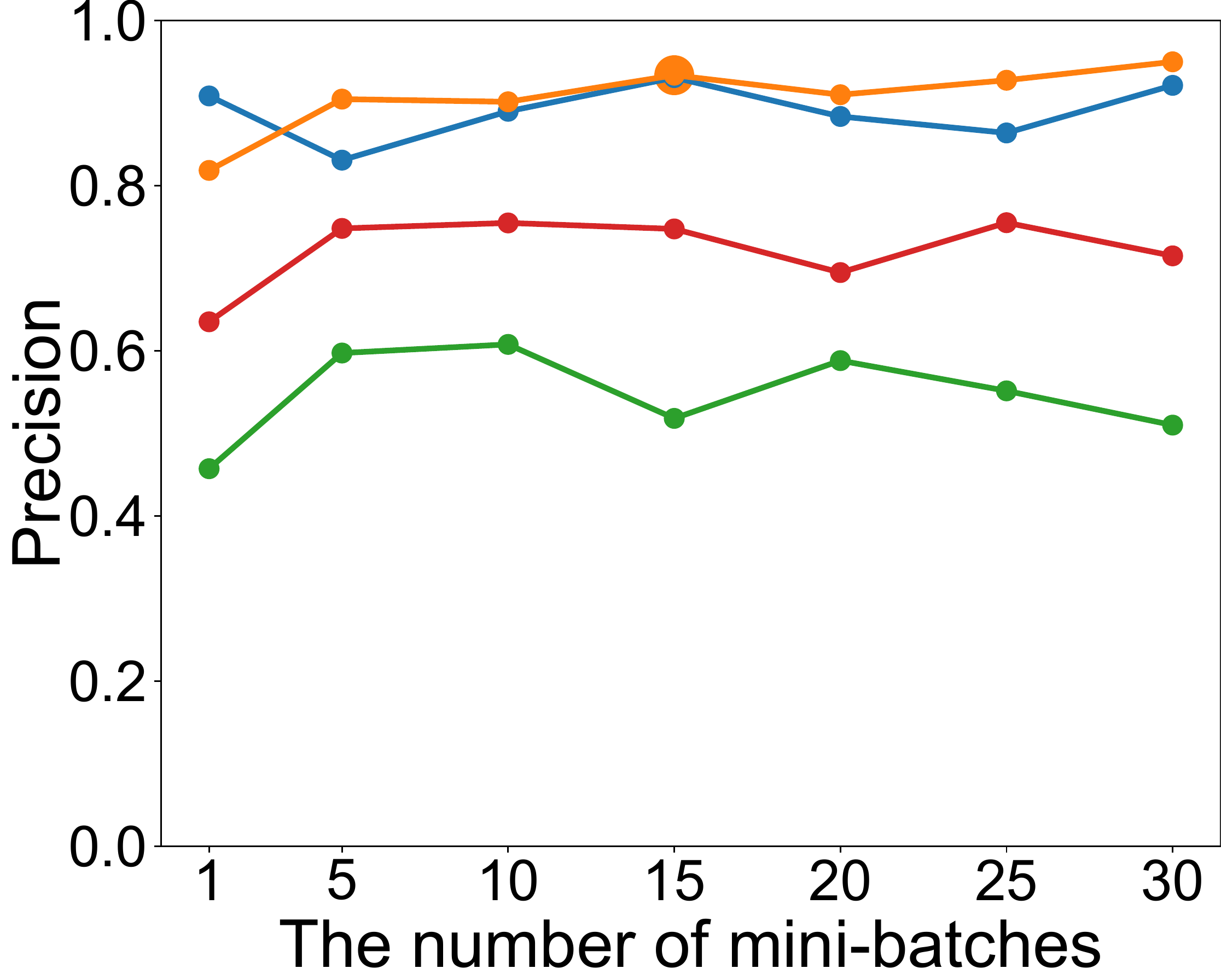}
        \subcaption{Precision of MPNN.}
    \end{minipage}
    \begin{minipage}[t]{0.32\linewidth}
        \centering
        \includegraphics[keepaspectratio, scale=0.22]{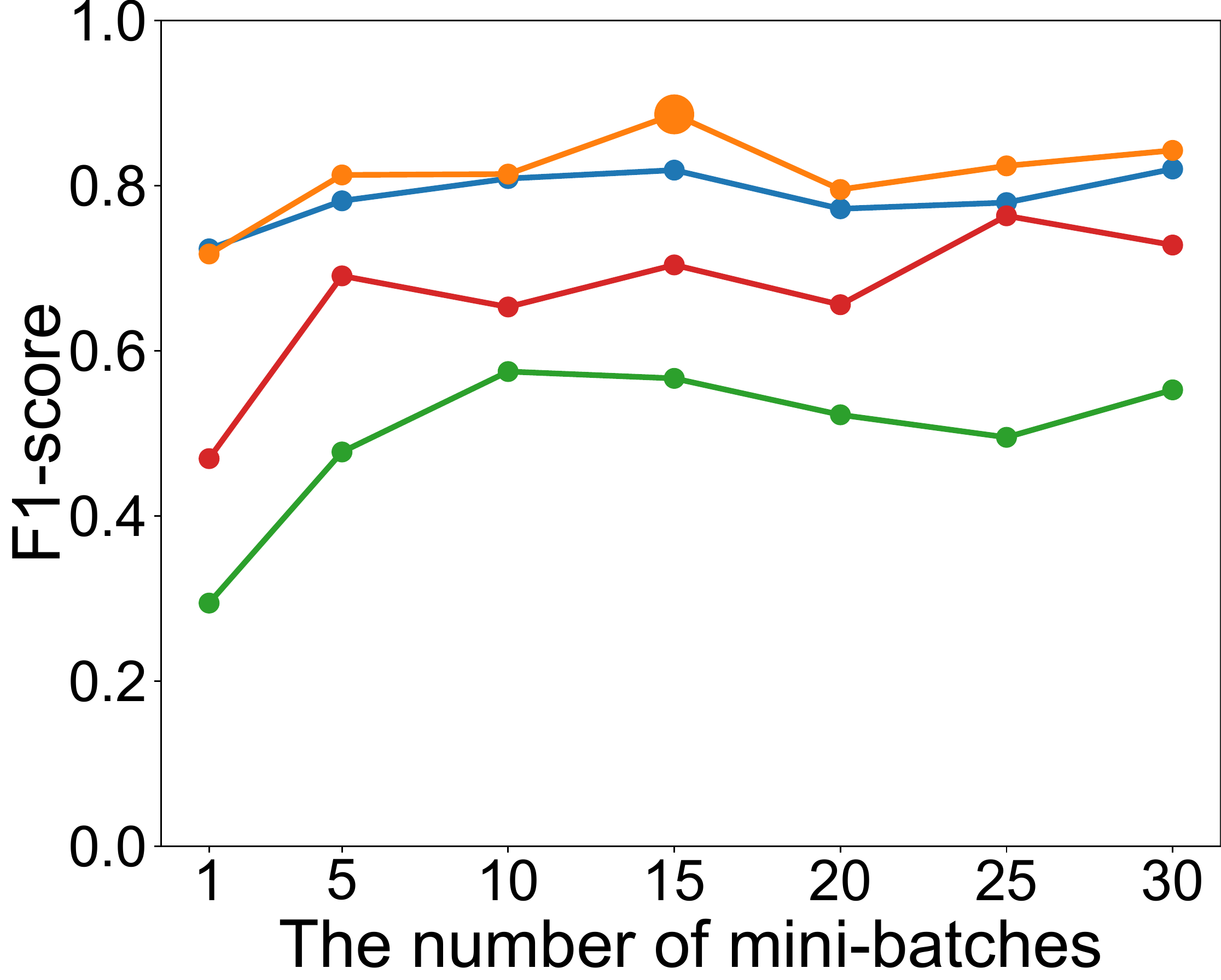}
        \subcaption{F1-score of MPNN.}
    \end{minipage}\\
    \begin{minipage}[t]{0.32\linewidth}
        \centering
        \includegraphics[keepaspectratio, scale=0.22]{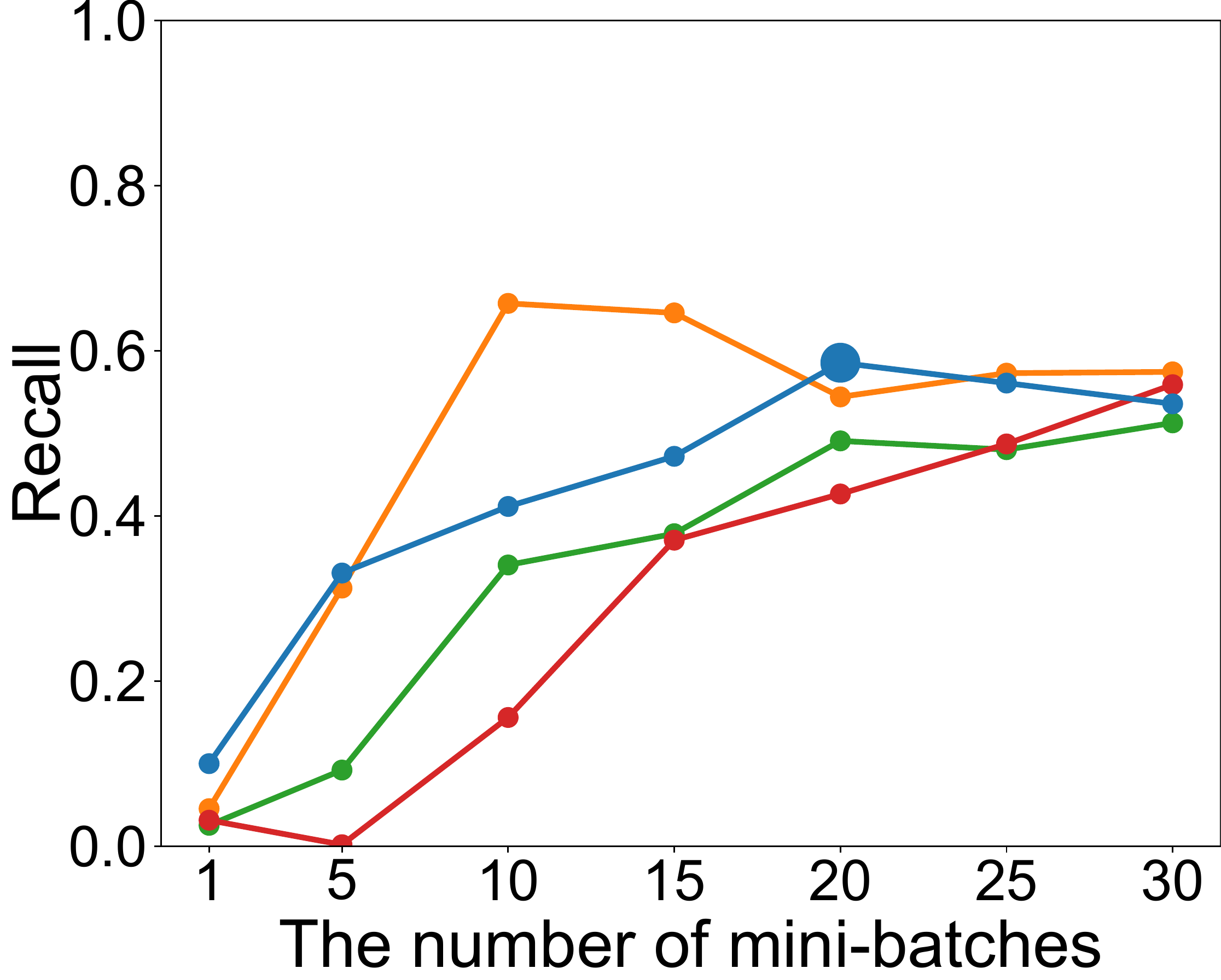}
        \subcaption{Recall of GIN.}
    \end{minipage}
    \begin{minipage}[t]{0.32\linewidth}
        \centering
        \includegraphics[keepaspectratio, scale=0.22]{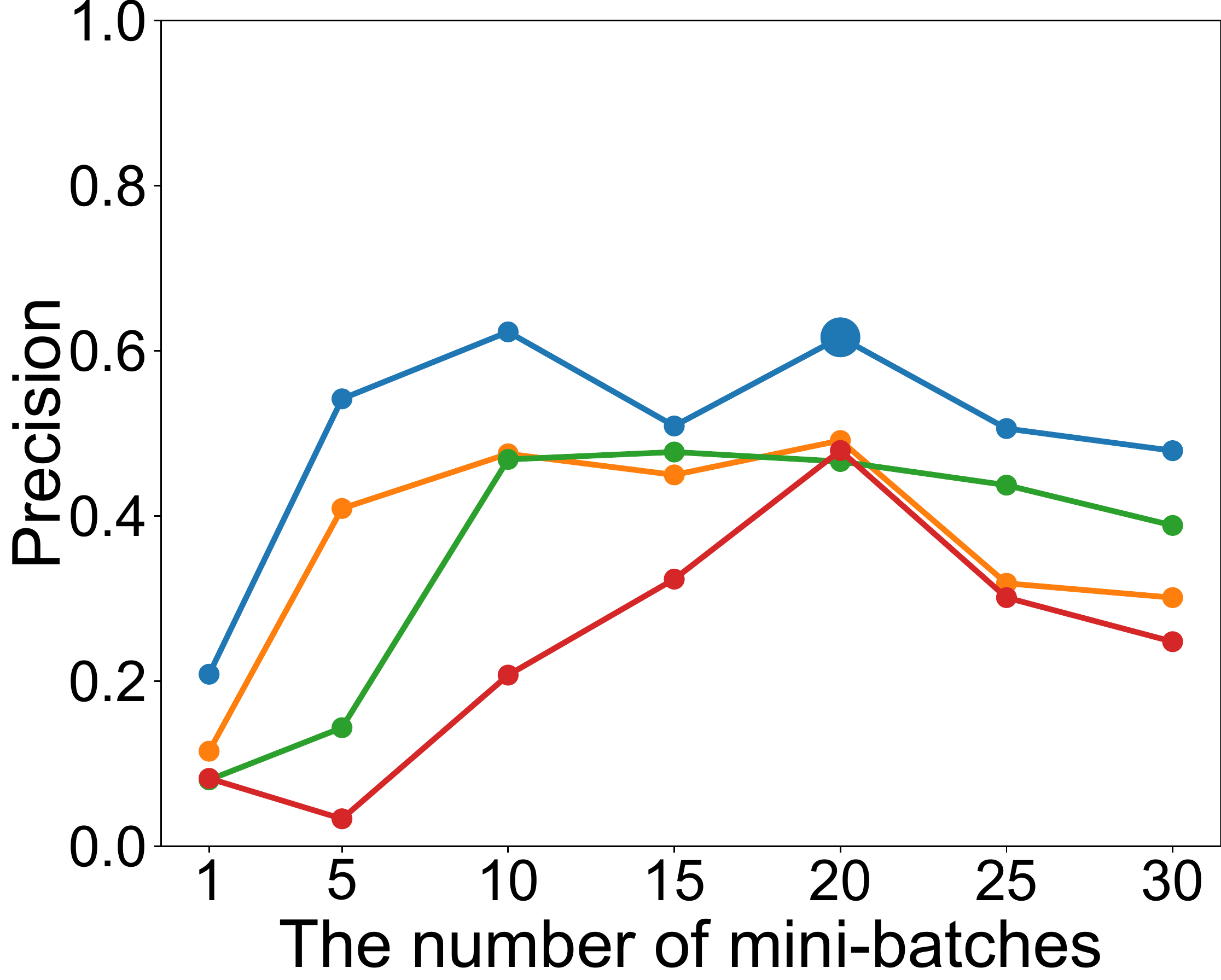}
        \subcaption{Precision of GIN.}
    \end{minipage}
    \begin{minipage}[t]{0.32\linewidth}
        \centering
        \includegraphics[keepaspectratio, scale=0.22]{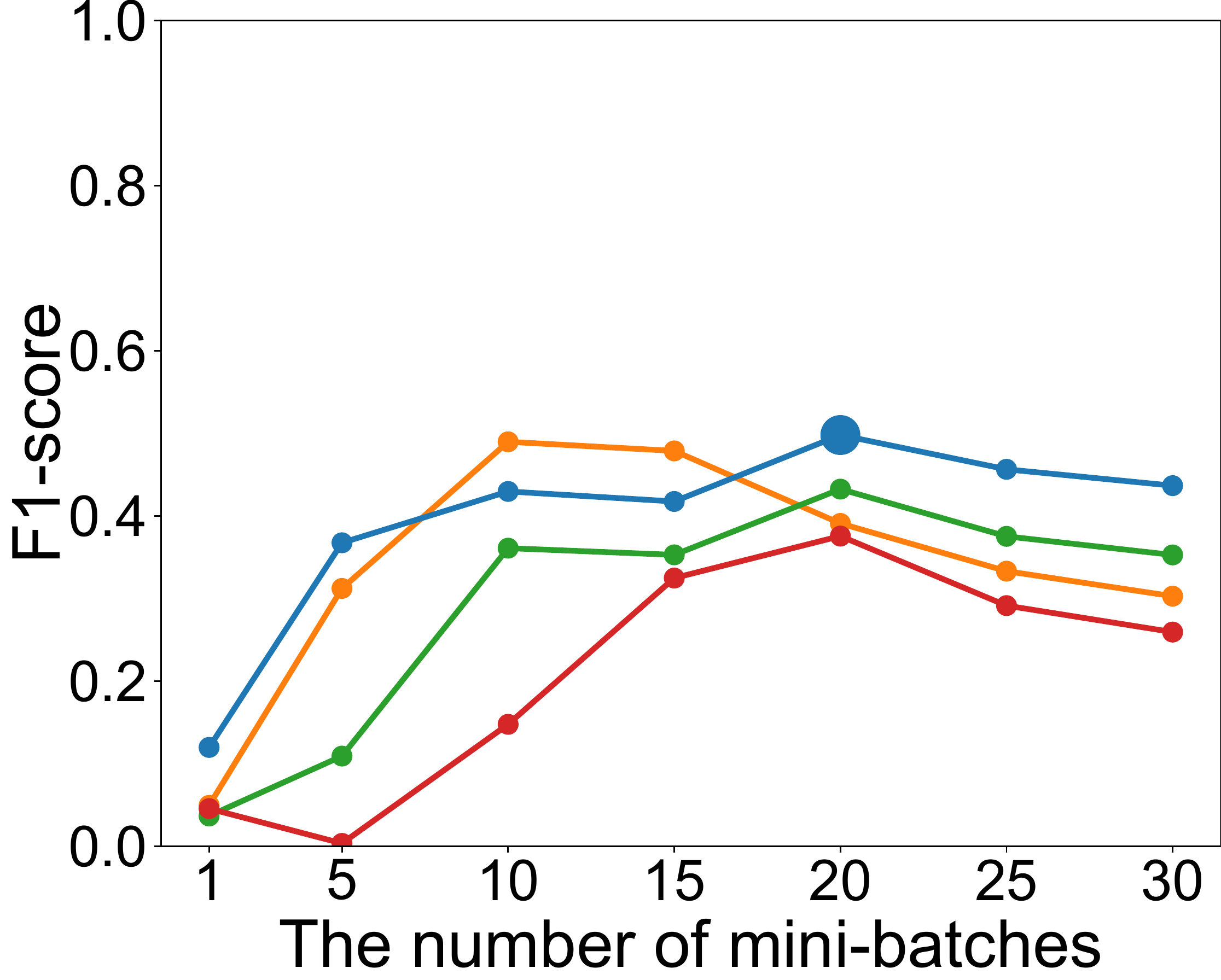}
        \subcaption{F1-score of GIN.}
    \end{minipage}\\
  \caption{Evaluation for the best parameters of the proposed method.}
  \label{fig:best-param-proposed}
\end{figure*}

\begin{figure*}[t]
\centering
    \begin{minipage}[t]{0.32\linewidth}
        \centering
        \includegraphics[keepaspectratio, scale=0.22]{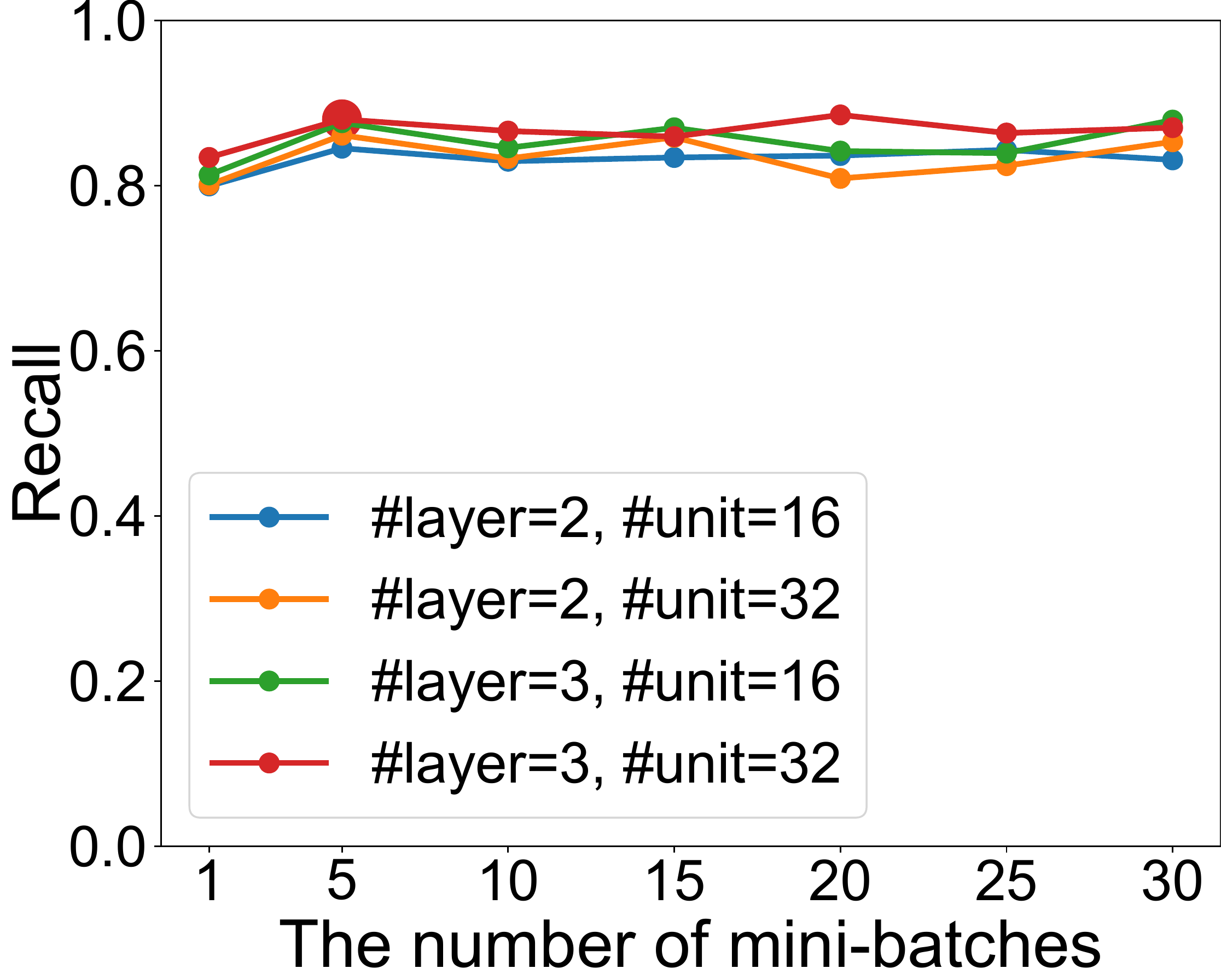}
        \subcaption{Recall of GAT.}
    \end{minipage}
    \begin{minipage}[t]{0.32\linewidth}
        \centering
        \includegraphics[keepaspectratio, scale=0.22]{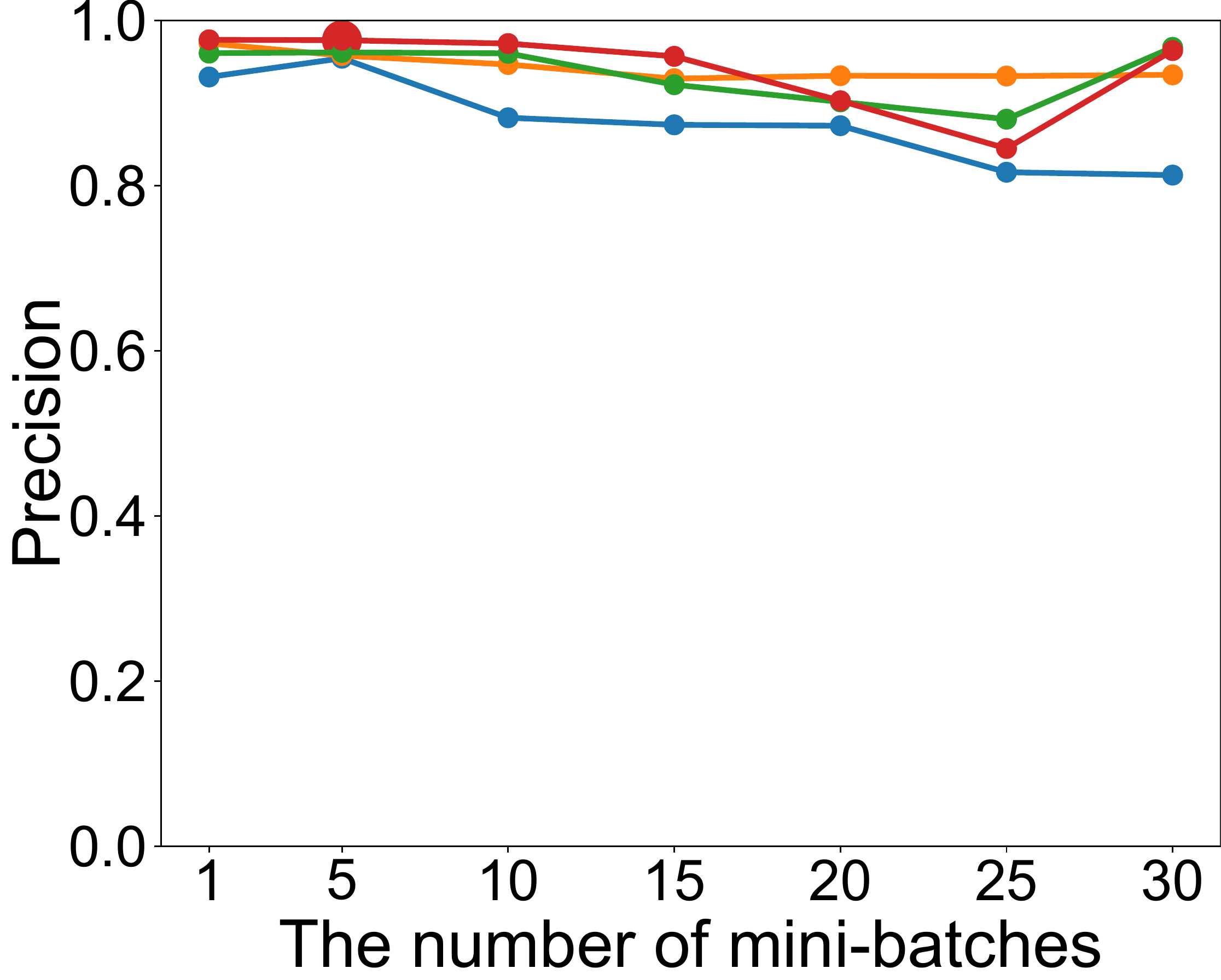}
        \subcaption{Precision of GAT.}
    \end{minipage}
    \begin{minipage}[t]{0.32\linewidth}
        \centering
        \includegraphics[keepaspectratio, scale=0.22]{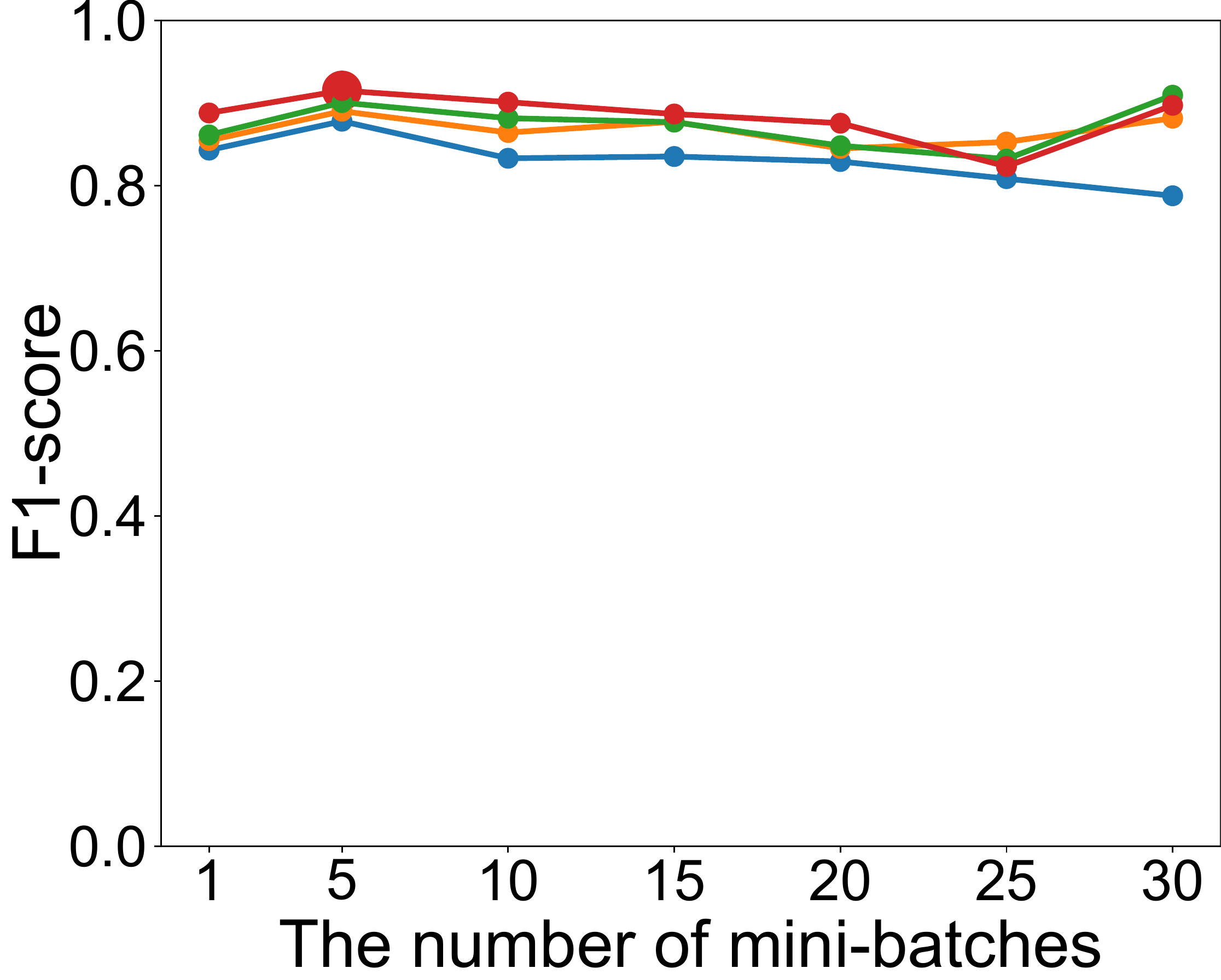}
        \subcaption{F1-score of GAT.}
    \end{minipage}\\
    \begin{minipage}[t]{0.32\linewidth}
        \centering
        \includegraphics[keepaspectratio, scale=0.22]{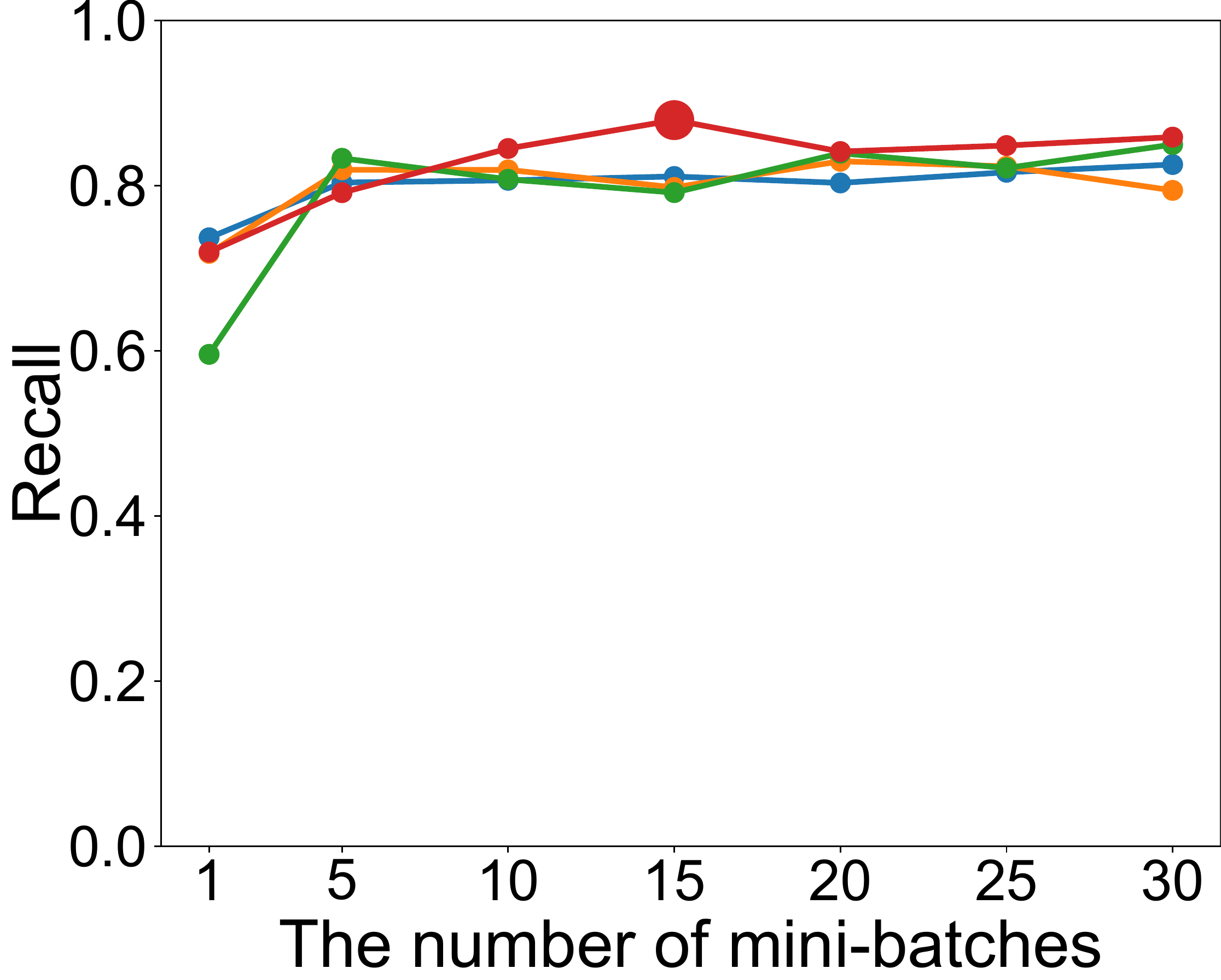}
        \subcaption{Recall of MPNN.}
    \end{minipage}
    \begin{minipage}[t]{0.32\linewidth}
        \centering
        \includegraphics[keepaspectratio, scale=0.22]{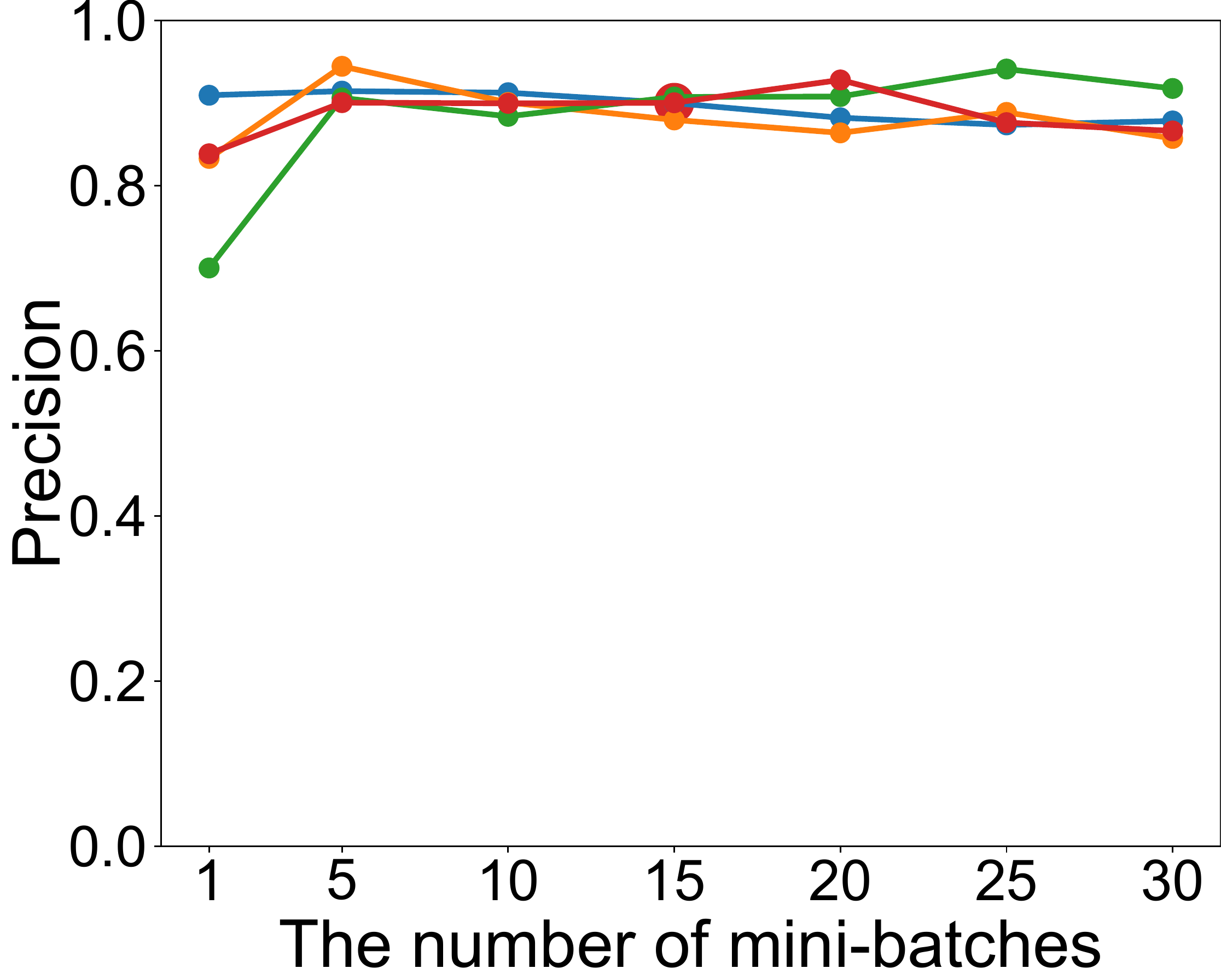}
        \subcaption{Precision of MPNN.}
    \end{minipage}
    \begin{minipage}[t]{0.32\linewidth}
        \centering
        \includegraphics[keepaspectratio, scale=0.22]{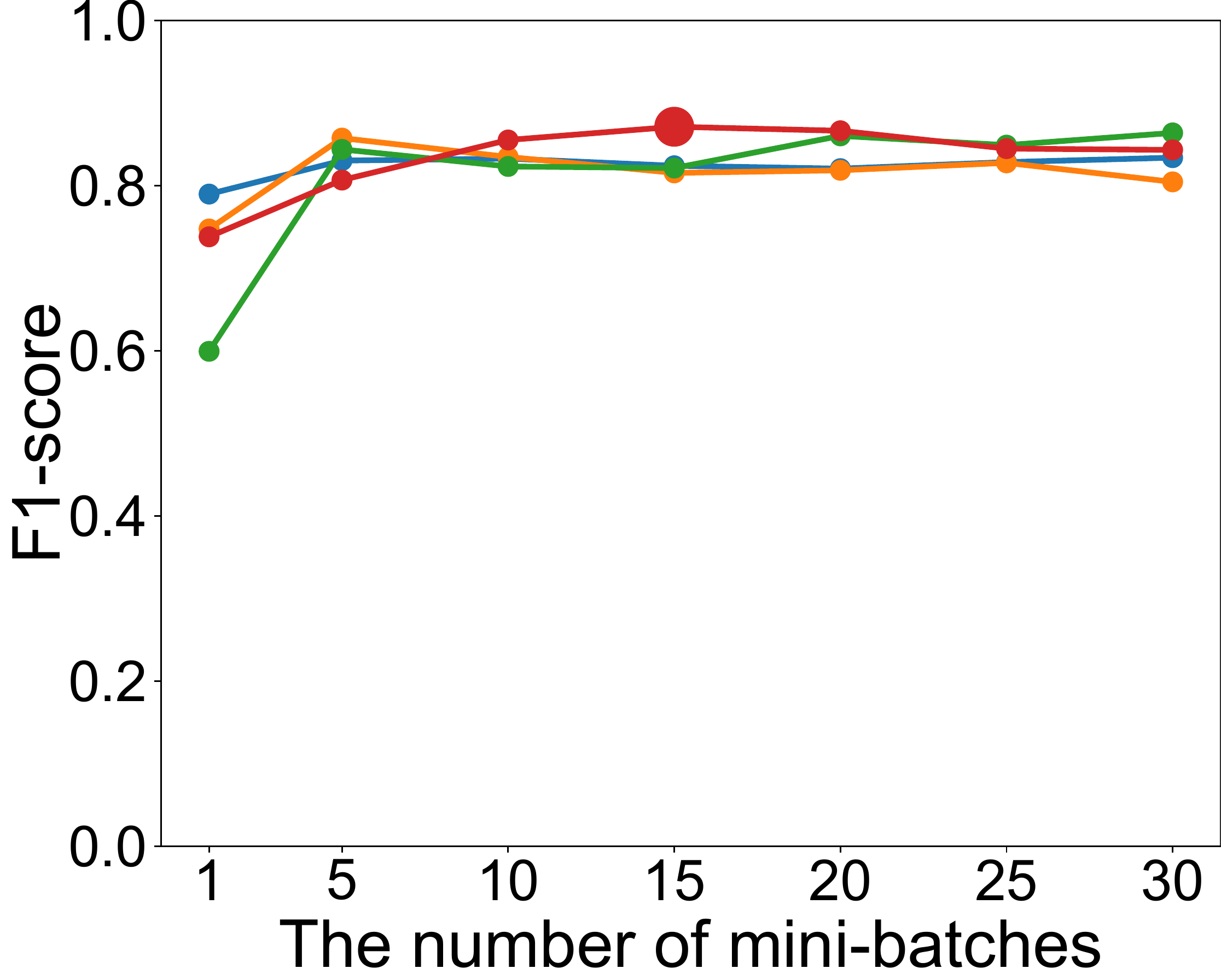}
        \subcaption{F1-score of MPNN.}
    \end{minipage}\\
    \begin{minipage}[t]{0.32\linewidth}
        \centering
        \includegraphics[keepaspectratio, scale=0.22]{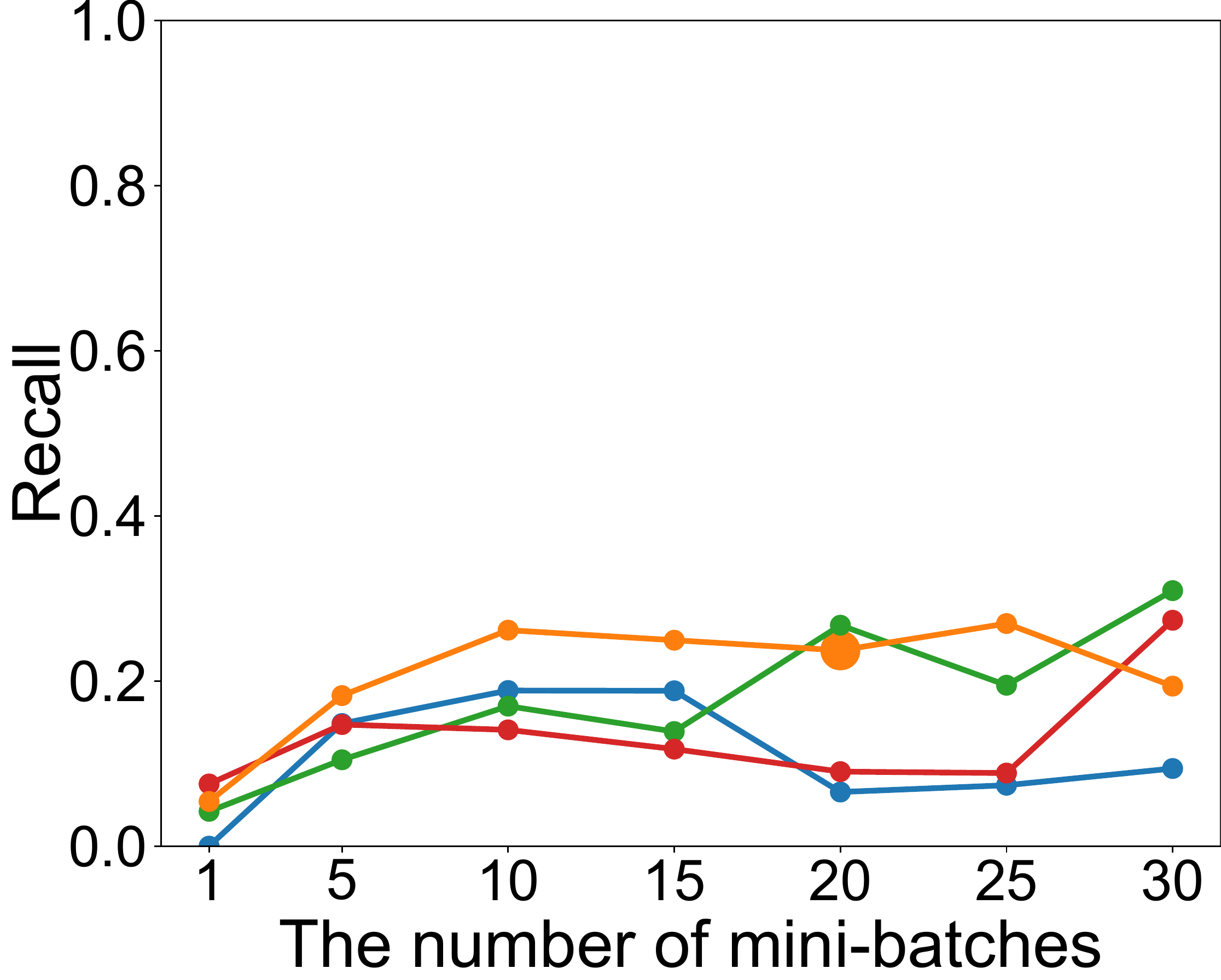}
        \subcaption{Recall of GIN.}
    \end{minipage}
    \begin{minipage}[t]{0.32\linewidth}
        \centering
        \includegraphics[keepaspectratio, scale=0.22]{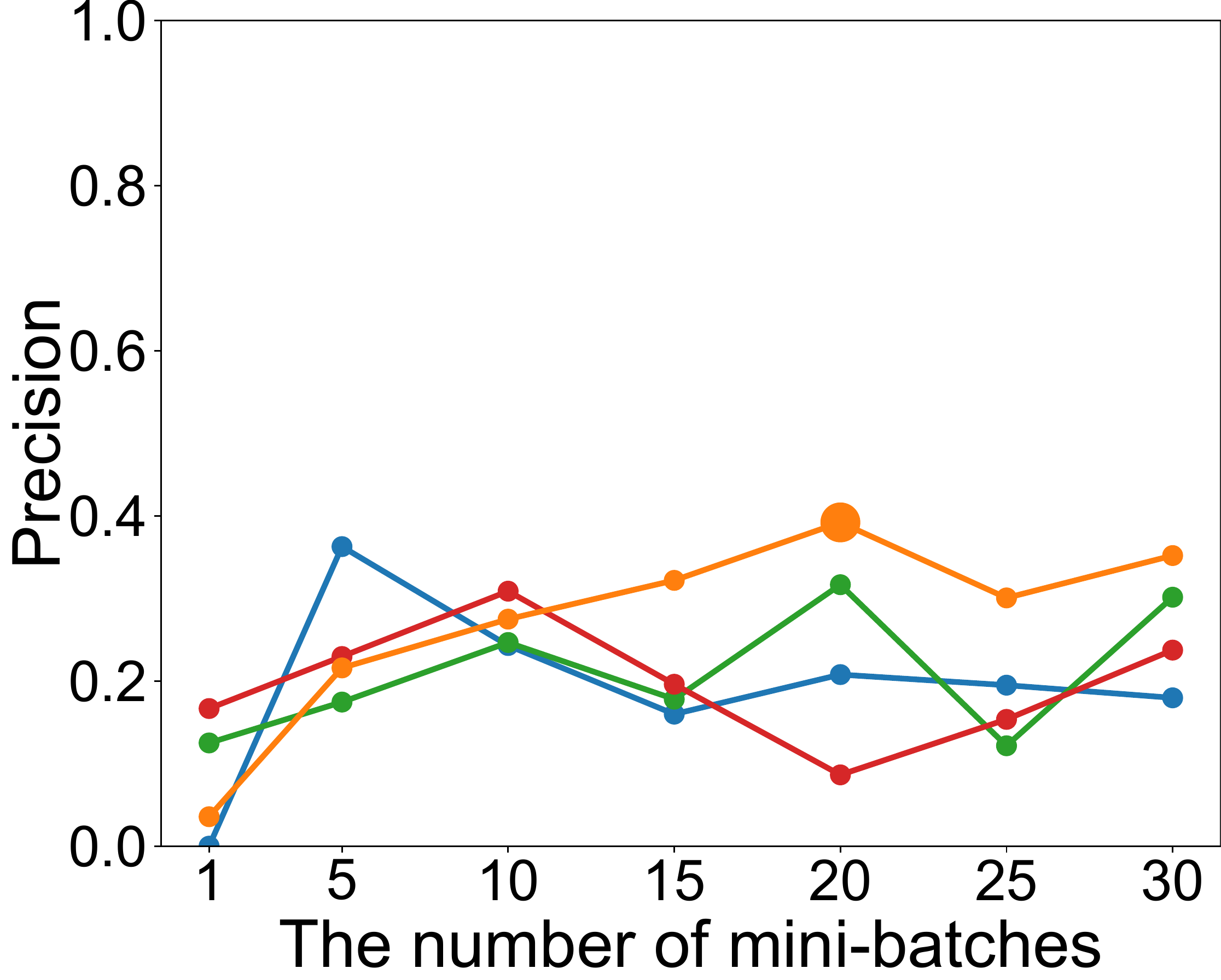}
        \subcaption{Precision of GIN.}
    \end{minipage}
    \begin{minipage}[t]{0.32\linewidth}
        \centering
        \includegraphics[keepaspectratio, scale=0.22]{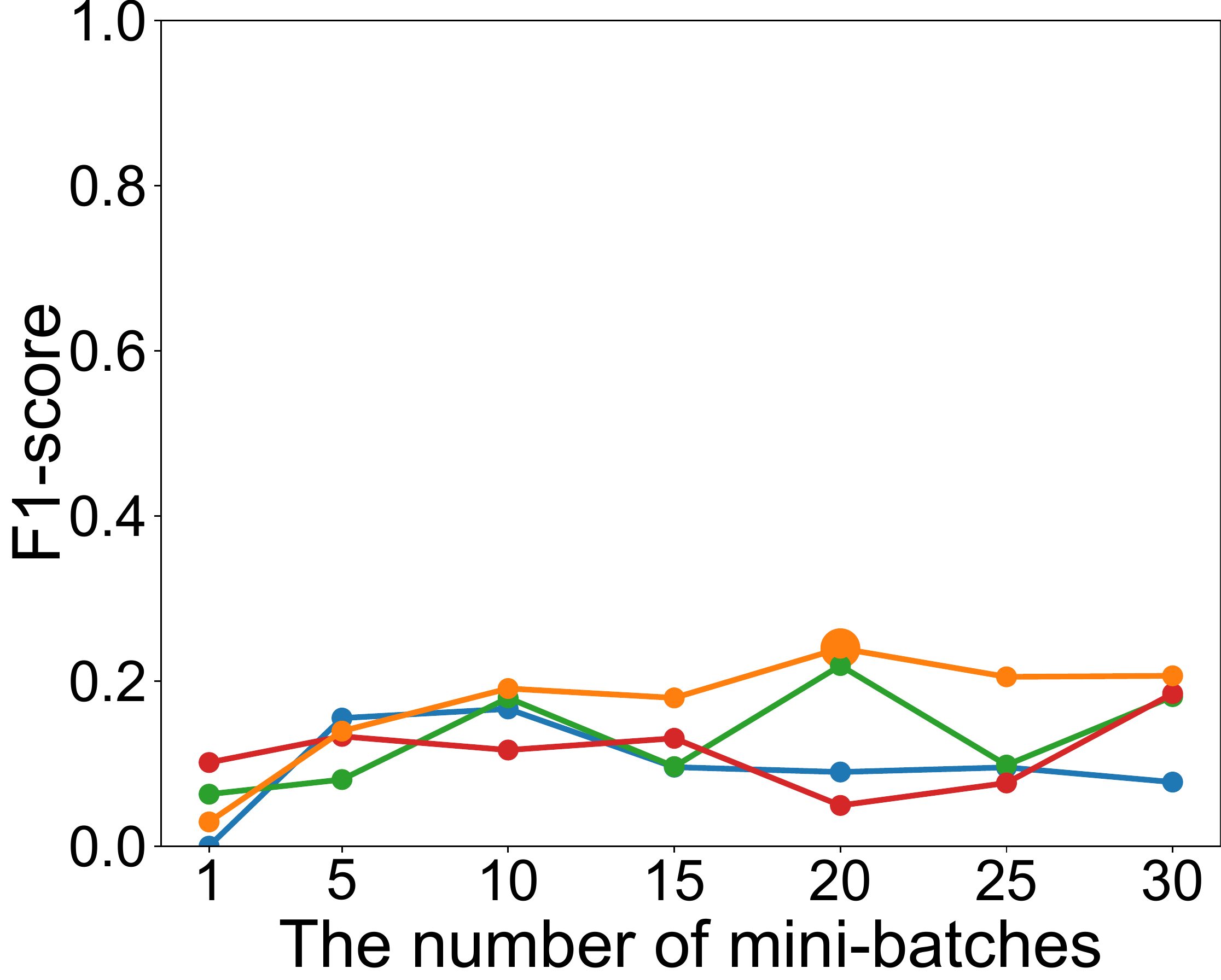}
        \subcaption{F1-score of GIN.}
    \end{minipage}\\
  \caption{Evaluation of the best parameters of the baseline method}
  \label{fig:best-param-baseline}
\end{figure*}

Figs.~\ref{fig:best-param-proposed} and \ref{fig:best-param-baseline} show all the evaluation results for searching for the best parameters of the proposed method and the baseline method, respectively.
In Figs.~\ref{fig:best-param-proposed} and \ref{fig:best-param-baseline}, we show the recall, precision, and F1-score results.
The x-axis shows the number of batches, and the y-axis shows the score of an evaluation metric in each plot.
The large-sized plots show the models that we have selected as the best ones.
\begin{table}[t]
    \centering
    \caption{Parameters of the experiment.}
    \label{tbl:parameters}
    \begin{tabular}{ccc}\hline
    \#batches                & \#layers  & \#units \\ \hline
    1, 5, 10, 15, 20, 25, 30 & 2, 3      & 16, 32 \\ \hline
    \end{tabular}
\end{table}

\section{Details of the Detection Results}
\label{apdx:best_parameter}

Tables~\ref{tbl:best_GAT}--\ref{tbl:base_best_GIN} show the detection results for each benchmark netlist using the GNN models with the best parameters obtained in Appendix~\ref{apdx:gnn_parameters}.
Tables~\ref{tbl:best_GAT}--\ref{tbl:best_GIN} (resp. Tables~\ref{tbl:base_best_GAT}--\ref{tbl:base_best_GIN}) are the results of the proposed method (resp. the baseline method).
We regard the class of Trojan nodes as `positive' and obtain the true positive~(TP), true negative~(TN), false positive~(FP), and false negative~(FN) based on the detection results from the GNN models.
The evaluation metrics are recall, precision, F1-score, and accuracy, which are expressed as follows: $\mathrm{Recall} = \mathrm{TP / (TP+FN)}$, $\mathrm{Precision} = \mathrm{TP / (TP+FP)}$, $\mathrm{F1\mathchar`-score} = \mathrm{(2 \cdot Recall \cdot Precision) / (Recall+Precision)}$, and $\mathrm{Accuracy} = \mathrm{(TP+TN) / (TP+FN+FP+TN)}$.
As shown in the tables, there are few FPs and FNs compared to the number of all the nodes in a netlist.
In particular, as shown in Table~\ref{tbl:best_GAT}, there are almost no FPs and few FNs for the GAT model, which is the best model among the three GNN models of the proposed method for the Trust-HUB benchmark netlists.

\begin{table}[t]
\centering
\caption{Detection results for the GAT model of the proposed method with the best parameter.}
\label{tbl:best_GAT}
\resizebox{\linewidth}{!}{%
\begin{tabular}{c|cccc|cccc} \hline
Netlist              & TN     & FP & FN  & TP  & Recall & Precision & F1-score & Accuracy \\ \hline \hline
RS232-T1000          & 289    & 0  & 0   & 13  & 1.000  & 1.000     & 1.000    & 1.000    \\
RS232-T1100          & 292    & 1  & 0   & 11  & 1.000  & 0.917     & 0.957    & 0.997    \\
RS232-T1200          & 295    & 1  & 0   & 10  & 1.000  & 0.909     & 0.952    & 0.997    \\
RS232-T1300          & 290    & 0  & 0   & 9   & 1.000  & 1.000     & 1.000    & 1.000    \\
RS232-T1400          & 289    & 1  & 0   & 12  & 1.000  & 0.923     & 0.960    & 0.997    \\
RS232-T1500          & 291    & 0  & 0   & 13  & 1.000  & 1.000     & 1.000    & 1.000    \\
RS232-T1600          & 290    & 0  & 4   & 9   & 0.692  & 1.000     & 0.818    & 0.987    \\
s15850-T100          & 2397   & 0  & 0   & 27  & 1.000  & 1.000     & 1.000    & 1.000    \\
s35932-T100          & 5967   & 0  & 3   & 12  & 0.800  & 1.000     & 0.889    & 0.999    \\
s35932-T200          & 5962   & 0  & 4   & 11  & 0.733  & 1.000     & 0.846    & 0.999    \\
s35932-T300          & 5972   & 3  & 5   & 21  & 0.808  & 0.875     & 0.840    & 0.999    \\
s38417-T100          & 5656   & 0  & 4   & 8   & 0.667  & 1.000     & 0.800    & 0.999    \\
s38417-T200          & 5656   & 0  & 0   & 15  & 1.000  & 1.000     & 1.000    & 1.000    \\
s38417-T300          & 5687   & 1  & 1   & 14  & 0.933  & 0.933     & 0.933    & 1.000    \\
s38584-T100          & 7063   & 0  & 6   & 3   & 0.333  & 1.000     & 0.500    & 0.999    \\
s38584-T200 & 7064   & 0  & 1   & 82  & 0.988  & 1.000     & 0.994    & 1.000    \\
s38584-T300 & 7064   & 0  & 177 & 554 & 0.758  & 1.000     & 0.862    & 0.977    \\
EthernetMAC10GE-T700 & 102453 & 0  & 0   & 13  & 1.000  & 1.000     & 1.000    & 1.000    \\
EthernetMAC10GE-T710 & 102452 & 0  & 0   & 13  & 1.000  & 1.000     & 1.000    & 1.000    \\
EthernetMAC10GE-T720 & 102453 & 0  & 1   & 12  & 0.923  & 1.000     & 0.960    & 1.000    \\
EthernetMAC10GE-T730 & 102453 & 0  & 0   & 13  & 1.000  & 1.000     & 1.000    & 1.000    \\
B19-T100             & 63170  & 0  & 0   & 83  & 1.000  & 1.000     & 1.000    & 1.000    \\
B19-T200             & 63161  & 9  & 0   & 83  & 1.000  & 0.902     & 0.949    & 1.000    \\
wb\_conmax-T100      & 23194  & 0  & 4   & 11  & 0.733  & 1.000     & 0.846    & 1.000    \\ \hline
Average              & -      & -  & -   & -   & 0.890  & 0.978     & 0.921    & 0.998   \\ \hline
\end{tabular}%
}
\end{table}

\begin{table}[t]
\centering
\caption{Detection results for the MPNN model of the proposed method with the best parameter.}
\label{tbl:best_MPNN}
\resizebox{\linewidth}{!}{%
\begin{tabular}{c|cccc|cccc}\hline
Netlist              & TN     & FP & FN & TP  & Recall & Precision & F1-score & Accuracy \\ \hline \hline
RS232-T1000          & 289    & 0  & 0  & 13  & 1.000  & 1.000     & 1.000    & 1.000    \\
RS232-T1100          & 293    & 0  & 0  & 11  & 1.000  & 1.000     & 1.000    & 1.000    \\
RS232-T1200          & 296    & 0  & 1  & 9   & 0.900  & 1.000     & 0.947    & 0.997    \\
RS232-T1300          & 288    & 2  & 0  & 9   & 1.000  & 0.818     & 0.900    & 0.993    \\
RS232-T1400          & 288    & 2  & 0  & 12  & 1.000  & 0.857     & 0.923    & 0.993    \\
RS232-T1500          & 290    & 1  & 0  & 13  & 1.000  & 0.929     & 0.963    & 0.997    \\
RS232-T1600          & 290    & 0  & 3  & 10  & 0.769  & 1.000     & 0.870    & 0.990    \\
s15850-T100          & 2387   & 10 & 2  & 25  & 0.926  & 0.714     & 0.806    & 0.995    \\
s35932-T100          & 5967   & 0  & 3  & 12  & 0.800  & 1.000     & 0.889    & 0.999    \\
s35932-T200          & 5961   & 1  & 2  & 13  & 0.867  & 0.929     & 0.897    & 0.999    \\
s35932-T300          & 5973   & 2  & 9  & 17  & 0.654  & 0.895     & 0.756    & 0.998    \\
s38417-T100          & 5656   & 0  & 6  & 6   & 0.500  & 1.000     & 0.667    & 0.999    \\
s38417-T200          & 5656   & 0  & 1  & 14  & 0.933  & 1.000     & 0.966    & 1.000    \\
s38417-T300          & 5687   & 1  & 4  & 11  & 0.733  & 0.917     & 0.815    & 0.999    \\
s38584-T100          & 7063   & 0  & 3  & 6   & 0.667  & 1.000     & 0.800    & 1.000    \\
s38584-T200 & 7060   & 4  & 0  & 83  & 1.000  & 0.954     & 0.976    & 0.999    \\
s38584-T300 & 7044   & 20 & 21 & 710 & 0.971  & 0.973     & 0.972    & 0.995    \\
EthernetMAC10GE-T700 & 102453 & 0  & 0  & 13  & 1.000  & 1.000     & 1.000    & 1.000    \\
EthernetMAC10GE-T710 & 102452 & 0  & 1  & 12  & 0.923  & 1.000     & 0.960    & 1.000    \\
EthernetMAC10GE-T720 & 102453 & 0  & 1  & 12  & 0.923  & 1.000     & 0.960    & 1.000    \\
EthernetMAC10GE-T730 & 102453 & 0  & 6  & 7   & 0.538  & 1.000     & 0.700    & 1.000    \\
B19-T100             & 63170  & 0  & 0  & 83  & 1.000  & 1.000     & 1.000    & 1.000    \\
B19-T200             & 63170  & 0  & 0  & 83  & 1.000  & 1.000     & 1.000    & 1.000    \\
wb\_conmax-T100      & 23180  & 14 & 5  & 10  & 0.667  & 0.417     & 0.513    & 0.999    \\ \hline
Average              & -      & -  & -  & -   & 0.865  & 0.933     & 0.887    & 0.998   \\ \hline
\end{tabular}%
}
\end{table}

\begin{table}[t]
\centering
\caption{Detection results for the GIN model of the proposed method with the best parameter.}
\label{tbl:best_GIN}
\resizebox{\linewidth}{!}{%
\begin{tabular}{c|cccc|cccc}\hline
Netlist  & TN     & FP   & FN & TP  & Recall & Precision & F1-score & Accuracy \\ \hline \hline
RS232-T1000          & 285    & 4    & 6  & 7   & 0.538  & 0.636     & 0.583    & 0.967    \\
RS232-T1100          & 289    & 4    & 6  & 5   & 0.455  & 0.556     & 0.500    & 0.967    \\
RS232-T1200          & 294    & 2    & 4  & 6   & 0.600  & 0.750     & 0.667    & 0.980    \\
RS232-T1300          & 287    & 3    & 9  & 0   & 0.000  & 0.000     & 0.000    & 0.960    \\
RS232-T1400          & 280    & 10   & 1  & 11  & 0.917  & 0.524     & 0.667    & 0.964    \\
RS232-T1500          & 289    & 2    & 7  & 6   & 0.462  & 0.750     & 0.571    & 0.970    \\
RS232-T1600          & 290    & 0    & 9  & 4   & 0.308  & 1.000     & 0.471    & 0.970    \\
s15850-T100          & 2389   & 8    & 1  & 26  & 0.963  & 0.765     & 0.852    & 0.996    \\
s35932-T100          & 5957   & 10   & 1  & 14  & 0.933  & 0.583     & 0.718    & 0.998    \\
s35932-T200          & 5961   & 1    & 13 & 2   & 0.133  & 0.667     & 0.222    & 0.998    \\
s35932-T300          & 5950   & 25   & 0  & 26  & 1.000  & 0.510     & 0.675    & 0.996    \\
s38417-T100          & 5656   & 0    & 5  & 7   & 0.583  & 1.000     & 0.737    & 0.999    \\
s38417-T200          & 5655   & 1    & 0  & 15  & 1.000  & 0.938     & 0.968    & 1.000    \\
s38417-T300          & 5663   & 25   & 2  & 13  & 0.867  & 0.342     & 0.491    & 0.995    \\
s38584-T100          & 7027   & 36   & 8  & 1   & 0.111  & 0.027     & 0.043    & 0.994    \\
s38584-T200 & 7063   & 1    & 1  & 82  & 0.988  & 0.988     & 0.988    & 1.000    \\
s38584-T300 & 7047   & 17   & 1  & 730 & 0.999  & 0.977     & 0.988    & 0.998    \\
EthernetMAC10GE-T700 & 102453 & 0    & 13 & 0   & 0.000  & 0.000     & 0.000    & 1.000    \\
EthernetMAC10GE-T710 & 102452 & 0    & 12 & 1   & 0.077  & 1.000     & 0.143    & 1.000    \\
EthernetMAC10GE-T720 & 102453 & 0    & 9  & 4   & 0.308  & 1.000     & 0.471    & 1.000    \\
EthernetMAC10GE-T730 & 102443 & 10   & 1  & 12  & 0.923  & 0.545     & 0.686    & 1.000    \\
B19-T100             & 62891  & 279  & 14 & 69  & 0.831  & 0.198     & 0.320    & 0.995    \\
B19-T200             & 60742  & 2428 & 1  & 82  & 0.988  & 0.033     & 0.063    & 0.962    \\
wb\_conmax-T100      & 23194  & 0    & 14 & 1   & 0.067  & 1.000     & 0.125    & 0.999    \\ \hline
Average              & -      & -    & -  & -   & 0.585  & 0.616     & 0.498    & 0.988   \\ \hline
\end{tabular}%
}
\end{table}

\begin{table}[t]
\centering
\caption{Detection results for the GAT model of the baseline method with the best parameter.}
\label{tbl:base_best_GAT}
\resizebox{\linewidth}{!}{%
\begin{tabular}{c|cccc|cccc}\hline
Netlist              & TN     & FP & FN & TP  & Recall & Precision & F1-score & Accuracy \\ \hline \hline
RS232-T1000          & 289    & 0  & 0  & 13  & 1.000  & 1.000     & 1.000    & 1.000    \\
RS232-T1100          & 293    & 0  & 0  & 11  & 1.000  & 1.000     & 1.000    & 1.000    \\
RS232-T1200          & 296    & 0  & 0  & 10  & 1.000  & 1.000     & 1.000    & 1.000    \\
RS232-T1300          & 290    & 0  & 0  & 9   & 1.000  & 1.000     & 1.000    & 1.000    \\
RS232-T1400          & 290    & 0  & 0  & 12  & 1.000  & 1.000     & 1.000    & 1.000    \\
RS232-T1500          & 289    & 2  & 0  & 13  & 1.000  & 0.867     & 0.929    & 0.993    \\
RS232-T1600          & 290    & 0  & 4  & 9   & 0.692  & 1.000     & 0.818    & 0.987    \\
s15850-T100          & 2388   & 9  & 2  & 25  & 0.926  & 0.735     & 0.820    & 0.995    \\
s35932-T100          & 5967   & 0  & 8  & 7   & 0.467  & 1.000     & 0.636    & 0.999    \\
s35932-T200          & 5962   & 0  & 5  & 10  & 0.667  & 1.000     & 0.800    & 0.999    \\
s35932-T300          & 5974   & 1  & 15 & 11  & 0.423  & 0.917     & 0.579    & 0.997    \\
s38417-T100          & 5656   & 0  & 1  & 11  & 0.917  & 1.000     & 0.957    & 1.000    \\
s38417-T200          & 5656   & 0  & 1  & 14  & 0.933  & 1.000     & 0.966    & 1.000    \\
s38417-T300          & 5688   & 0  & 2  & 13  & 0.867  & 1.000     & 0.929    & 1.000    \\
s38584-T100          & 7063   & 0  & 3  & 6   & 0.667  & 1.000     & 0.800    & 1.000    \\
s38584-T200 & 7062   & 2  & 1  & 82  & 0.988  & 0.976     & 0.982    & 1.000    \\
s38584-T300 & 7061   & 3  & 82 & 649 & 0.888  & 0.995     & 0.939    & 0.989    \\
EthernetMAC10GE-T700 & 102453 & 0  & 0  & 13  & 1.000  & 1.000     & 1.000    & 1.000    \\
EthernetMAC10GE-T710 & 102452 & 0  & 1  & 12  & 0.923  & 1.000     & 0.960    & 1.000    \\
EthernetMAC10GE-T720 & 102453 & 0  & 1  & 12  & 0.923  & 1.000     & 0.960    & 1.000    \\
EthernetMAC10GE-T730 & 102453 & 0  & 1  & 12  & 0.923  & 1.000     & 0.960    & 1.000    \\
B19-T100             & 63170  & 0  & 0  & 83  & 1.000  & 1.000     & 1.000    & 1.000    \\
B19-T200             & 63170  & 0  & 1  & 82  & 0.988  & 1.000     & 0.994    & 1.000    \\
wb\_conmax-T100      & 23193  & 1  & 1  & 14  & 0.933  & 0.933     & 0.933    & 1.000    \\ \hline
Average              & -      & -  & -  & -   & 0.880  & 0.976     & 0.915    & 0.998   \\ \hline
\end{tabular}%
}
\end{table}

\begin{table}[t]
\centering
\caption{Detection results for the MPNN model of the baseline method with the best parameter.}
\label{tbl:base_best_MPNN}
\resizebox{\linewidth}{!}{%
\begin{tabular}{c|cccc|cccc}\hline
Netlist              & TN     & FP & FN  & TP  & Recall & Precision & F1-score & Accuracy \\ \hline \hline
RS232-T1000          & 289    & 0  & 0   & 13  & 1.000  & 1.000     & 1.000    & 1.000    \\
RS232-T1100          & 293    & 0  & 0   & 11  & 1.000  & 1.000     & 1.000    & 1.000    \\
RS232-T1200          & 296    & 0  & 0   & 10  & 1.000  & 1.000     & 1.000    & 1.000    \\
RS232-T1300          & 290    & 0  & 0   & 9   & 1.000  & 1.000     & 1.000    & 1.000    \\
RS232-T1400          & 290    & 0  & 0   & 12  & 1.000  & 1.000     & 1.000    & 1.000    \\
RS232-T1500          & 291    & 0  & 0   & 13  & 1.000  & 1.000     & 1.000    & 1.000    \\
RS232-T1600          & 290    & 0  & 4   & 9   & 0.692  & 1.000     & 0.818    & 0.987    \\
s15850-T100          & 2368   & 29 & 1   & 26  & 0.963  & 0.473     & 0.634    & 0.988    \\
s35932-T100          & 5967   & 0  & 5   & 10  & 0.667  & 1.000     & 0.800    & 0.999    \\
s35932-T200          & 5962   & 0  & 7   & 8   & 0.533  & 1.000     & 0.696    & 0.999    \\
s35932-T300          & 5974   & 1  & 12  & 14  & 0.538  & 0.933     & 0.683    & 0.998    \\
s38417-T100          & 5656   & 0  & 2   & 10  & 0.833  & 1.000     & 0.909    & 1.000    \\
s38417-T200          & 5656   & 0  & 1   & 14  & 0.933  & 1.000     & 0.966    & 1.000    \\
s38417-T300          & 5662   & 26 & 2   & 13  & 0.867  & 0.333     & 0.481    & 0.995    \\
s38584-T100          & 7062   & 1  & 1   & 8   & 0.889  & 0.889     & 0.889    & 1.000    \\
s38584-T200 & 7059   & 5  & 1   & 82  & 0.988  & 0.943     & 0.965    & 0.999    \\
s38584-T300 & 7059   & 5  & 118 & 613 & 0.839  & 0.992     & 0.909    & 0.984    \\
EthernetMAC10GE-T700 & 102451 & 2  & 2   & 11  & 0.846  & 0.846     & 0.846    & 1.000    \\
EthernetMAC10GE-T710 & 102452 & 0  & 1   & 12  & 0.923  & 1.000     & 0.960    & 1.000    \\
EthernetMAC10GE-T720 & 102451 & 2  & 1   & 12  & 0.923  & 0.857     & 0.889    & 1.000    \\
EthernetMAC10GE-T730 & 102453 & 0  & 0   & 13  & 1.000  & 1.000     & 1.000    & 1.000    \\
B19-T100             & 63170  & 0  & 0   & 83  & 1.000  & 1.000     & 1.000    & 1.000    \\
B19-T200             & 63166  & 4  & 0   & 83  & 1.000  & 0.954     & 0.976    & 1.000    \\
wb\_conmax-T100      & 23178  & 16 & 5   & 10  & 0.667  & 0.385     & 0.488    & 0.999    \\ \hline
Average              & -      & -  & -   & -   & 0.879  & 0.900     & 0.871    & 0.998   \\ \hline
\end{tabular}%
}
\end{table}

\begin{table}[t]
\centering
\caption{Detection results for the GIN model of the baseline method with the best parameter.}
\label{tbl:base_best_GIN}
\resizebox{\linewidth}{!}{%
\begin{tabular}{c|cccc|cccc}\hline
Netlist              & TN     & FP  & FN  & TP  & Recall & Precision & F1-score & Accuracy \\ \hline \hline
RS232-T1000          & 287    & 2   & 1   & 12  & 0.923  & 0.857     & 0.889    & 0.990    \\
RS232-T1100          & 292    & 1   & 11  & 0   & 0.000  & 0.000     & 0.000    & 0.961    \\
RS232-T1200          & 296    & 0   & 10  & 0   & 0.000  & 0.000     & 0.000    & 0.967    \\
RS232-T1300          & 290    & 0   & 9   & 0   & 0.000  & 0.000     & 0.000    & 0.970    \\
RS232-T1400          & 289    & 1   & 9   & 3   & 0.250  & 0.750     & 0.375    & 0.967    \\
RS232-T1500          & 291    & 0   & 13  & 0   & 0.000  & 0.000     & 0.000    & 0.957    \\
RS232-T1600          & 290    & 0   & 12  & 1   & 0.077  & 1.000     & 0.143    & 0.960    \\
s15850-T100          & 2397   & 0   & 27  & 0   & 0.000  & 0.000     & 0.000    & 0.989    \\
s35932-T100          & 5967   & 0   & 4   & 11  & 0.733  & 1.000     & 0.846    & 0.999    \\
s35932-T200          & 5962   & 0   & 13  & 2   & 0.133  & 1.000     & 0.235    & 0.998    \\
s35932-T300          & 5972   & 3   & 23  & 3   & 0.115  & 0.500     & 0.188    & 0.996    \\
s38417-T100          & 5656   & 0   & 12  & 0   & 0.000  & 0.000     & 0.000    & 0.998    \\
s38417-T200          & 5652   & 4   & 14  & 1   & 0.067  & 0.200     & 0.100    & 0.997    \\
s38417-T300          & 5678   & 10  & 14  & 1   & 0.067  & 0.091     & 0.077    & 0.996    \\
s38584-T100          & 7058   & 5   & 5   & 4   & 0.444  & 0.444     & 0.444    & 0.999    \\
s38584-T200 & 7028   & 36  & 7   & 76  & 0.916  & 0.679     & 0.779    & 0.994    \\
s38584-T300 & 7023   & 41  & 386 & 345 & 0.472  & 0.894     & 0.618    & 0.945    \\
EthernetMAC10GE-T700 & 102429 & 24  & 8   & 5   & 0.385  & 0.172     & 0.238    & 1.000    \\
EthernetMAC10GE-T710 & 102431 & 21  & 11  & 2   & 0.154  & 0.087     & 0.111    & 1.000    \\
EthernetMAC10GE-T720 & 102428 & 25  & 8   & 5   & 0.385  & 0.167     & 0.233    & 1.000    \\
EthernetMAC10GE-T730 & 102451 & 2   & 11  & 2   & 0.154  & 0.500     & 0.235    & 1.000    \\
B19-T100             & 63170  & 0   & 83  & 0   & 0.000  & 0.000     & 0.000    & 0.999    \\
B19-T200             & 62777  & 393 & 54  & 29  & 0.349  & 0.069     & 0.115    & 0.993    \\
wb\_conmax-T100      & 23194  & 0   & 14  & 1   & 0.067  & 1.000     & 0.125    & 0.999    \\ \hline
Average              & -      & -   & -   & -   & 0.237  & 0.392     & 0.240    & 0.986   \\ \hline
\end{tabular}%
}
\end{table}

\section{Random HT Generation}
\label{apdx:random_ht_generation}

For the experiment in Section~\ref{subsec:6_3_unknown_features}, we implement a method to randomly generate HT-infested circuits inspired by \cite{conf/date/CruzHMB18}.
In our implementation, we consider that the HT circuit consists of a trigger part and payload part.
We prepare several templates of the parts, as shown in Table~\ref{tbl:ht-generation-templates}, as candidates of HT circuits.

\begin{table}[t]
    \centering
    \caption{Templates used in random HT generation}
    \label{tbl:ht-generation-templates}
    \begin{tabular}{l|l} \hline
        Part & Template \\ \hline
        Trigger & Combinatorial, Sequential \\ 
        Payload & Denial of service, Information leakage, Power consuming \\ \hline
    \end{tabular}
\end{table}

In the experiments, we select five normal circuits from the Trust-HUB benchmark, and generate 20 samples for each normal circuit.
The template is written in RTL, and we synthesize it with Synopsis Design Compiler using the Synopsys SAED 90nm standard cell library.
When generating a sample, the templates in Table~\ref{tbl:ht-generation-templates} are randomly chosen as a trigger and payload part with the internal parameters of the parts randomly configured.
A sample is composed by connecting the configured parts and a normal circuit.
